# Methodological Framework for Determining the Land Eligibility of Renewable Energy Sources

David Severin Ryberg[a], Martin Robinius[a], Detlef Stolten[a,b]

[a] Institute of Electrochemical Process Engineering (IEK-3), Forschungszentrum Jülich GmbH, Wilhelm-Johnen-Str., 52428 Jülich, Germany
[b] Chair for Fuel Cells, RWTH Aachen University, c/o Institute of Electrochemical Process Engineering (IEK-3), Forschungszentrum Jülich GmbH, Wilhelm-Johnen-Str., 52428 Jülich, Germany

*Abstract*—**The quantity and distribution of land which is eligible for renewable energy sources is fundamental to the role these technologies will play in future energy systems. As it stands, however, the current state of land eligibility investigation is found to be insufficient to meet the demands of the future energy modelling community. Three key areas are identified as the predominate causes of this; inconsistent criteria definitions, inconsistent or unclear methodologies, and inconsistent dataset usage. To combat these issues, a land eligibility framework is developed and described in detail. The validity of this framework is then shown via the recreation of land eligibility results found in the literature, showing strong agreement in the majority of cases. Following this, the framework is used to perform an evaluation of land eligibility criteria within the European context whereby the relative importance of commonly considered criteria are compared.**

## 1. INTRODUCTION

As many world economies aim to meet emission reduction targets, countries will need to carefully consider the options available to them when choosing how to develop their energy systems. Choosing a particular developmental pathway is a challenging endeavor, however, given the uncertainties of future climate impacts and evolving sociotechnical landscapes. Therefore, an effort must be made to explore as much as possible the different pathway options available and future scenarios that might arise. In this regard, progress is being made in the form of energy system design models and similar analyses which serve to evaluate these pathways [2-4]. However, the situation is complicated by the fact that the pathways that various countries choose are not independent of one another [5]. For this reason, a globally-applicable solution can only be reached via communication and cooperation between the many research groups and organizations performing these evaluations, as well as consistency between their approaches, such that their results can be compared against each other's.

Judging from recent trends renewable energy sources (RES) will certainly play a significant role in the energy mix of these evaluated developmental pathways [6, 7]. Amongst other technologies, this will likely include on- and off-shore wind turbines, photovoltaic (PV) arrays, concentrated solar power (CSP) parks and biomass processing plants. Well known issues that these technologies entail, such as their intermittent [8, 9] and spatially-dependent [10] power production, have been the focus of intense research for many decades. Nevertheless, many uncertainties and unanswered questions persist that prevent the guarantee of successful implementation of large-scale RES technologies into future energy systems. Of these uncertainties, the influence of sociotechnical criteria, such as natural conservation, disruptions to local populations, and unfit terrain on the distribution of RES technologies across a region is outstanding. When small or otherwise uniform study regions exhibit little variance in their spatial characteristics, the consequences of a variable distribution can be largely ignored or simplified, yet as evaluations progress towards larger spatial scope, this variability quickly becomes a crucial quality to consider [11]. One of the main reasons this issue remains outstanding, however, is that a region's response to these sociotechnical criteria are dependent not only on the technology being considered, but can vary significantly between one region and another [12]. Moreover, even when investigating a particular technology within a given region, the region's response to these criteria will likely change over time alongside evolving social preferences and technological advances [13]. Therefore, it is apparent that when evaluating these developmental pathways in broad spatial contexts, the proper treatment of RES components is dependent on a methodological application of the spatially-sensitive sociotechnical criteria governing where these technologies can be installed.

The application of sociotechnical criteria is inherently a geospatial question, which has, in fact, received significant attention from the research community [14-16]. One simple avenue in which these criteria affect RES distribution is conveyed by the concept of land eligibility (LE); the binary conclusion of whether or not a plot of land is eligible for RES installation based on the relevant criteria. LE has been described by Iqbal [3] as one of the typical inputs in the generic energy resource allocation problem, which includes energy system design and many other LE-dependent problems as well. Alongside LE, sociotechnical criteria are also commonly employed in Multi Criterion Decision Management (MCDM) analyses which include, amongst other examples, the well-known Analytic Hierarchy Process (AHP) [17]. In this context, MCDM analyses aim to estimate the relative likelihood of actually utilizing any one location. Moreover, MCDM is fundamentally different from LE despite primarily relying on the same criteria. Examples of LE analyses in the literature are common [1, 18-22] and many studies perform both an LE and MCDM analysis alongside one another [23-29].

However, despite this attention from the community, inconsistencies between studies have prevented a collective understanding of how sociotechnical criteria influence RES distribution [14]. As a result, the current situation remains insufficient to meet the requirements of researchers and policymakers who wish to make effective evaluations of the pathway development options open to them. Therefore, this work aims to promote awareness of the issues plaguing this field and to then present options that can help alleviate some of these issues. To do this, the application of LE analysis is heavily emphasized in this work due to its relative simplicity, although many of the issues discussed here relate, or even apply directly, to more complex analyses as well (such as MCDM).

The paper is structured as follows: First, a general overview of Geographic Information Systems (GIS) utilization in the literature is provided, summarizing general LE approaches and discussing why the current state is ill-suited for application on the broad scale of energy system evaluation. Following this, an LE framework is proposed that allows for consistent LE applications in any regional or technological context. After describing this framework, a validation is performed showing that its use can closely match the results of other LE evaluations and is therefore reliable. Finally, the described framework is utilized to investigate interactions between common sociotechnical criteria in the context of Europe in order to help build understanding of their regional relevance and relative importance.

## 2. LAND ELIGIBILITY

For the purposes of this discussion, the land eligibility (LE) of a particular RES technology is defined as the binary conclusion dictating whether the technology in question is allowed to be placed at a particular location. A location, in this sense, can be thought of as an area of land somewhere on the Earth's surface that is small enough to be considered as an aggregate. LE analyses are not generally concerned with the eligibility of a single location, however, but instead investigate a set of locations that in total comprise a region. The rules that lead to a location being deemed available or ineligible are understood from a set of exclusion constraints. Furthermore, exclusion constraint sets are typically unique for different RES technologies and when applied in different regions.

For the sake of clarity, the following terminology is defined before the discussion of LE is continued. *Criteria* refer to the attributes of land and can be defined in any way, as long as an unambiguous value can be assigned to every location. Common criteria definitions seen in the literature include "the distance to the nearest road" (for examples of this use, see [30-32]), "the average terrain slope" (ex. [33-35]), and "the predominant land cover" (forests [36], grasslands [37], urban areas [38-40], etc.). A criterion gives a value for each location and from this an *exclusion constraint* dictates which values result in a location's ineligibility. Exclusion constraints generally take the form of a threshold, eliminating all locations with criterion values either above or below the threshold depending on the nature of the constraint. However, they can also take the form of an acceptable range or an acceptable set of values as well. In comparison, *decision factors* are rules that neither guarantee nor exclude the siting of an RES technology at a given location, but nevertheless play a role in determining how likely such a placement would be. Decision factors are also defined from the same criteria as exclusion constraints, but since they do not dictate eligibility they are not used within LE analyses; they are, however, used in MCDM analyses. As an example, a policy prohibiting the installation of PV panels within designated protected biospheres (since the modules are likely to be incompatible with the flora and fauna found in the area) would be considered an exclusion constraint. Meanwhile, the average annual irradiance at some location would naturally impact the final likelihood of placing a panel at that location, but is more aptly treated as a decision factor since all locations will still receive some radiation.

### 2.1. Role of Land Eligibility Analyses

Examples of LE analyses in the literature are common and, among them, are notable examples of broad-context investigations. The European Environmental Agency (EEA) [41], for one, investigated the LE of onshore wind turbines in the European Union where the avoidance of protected areas was the only exclusion constraint. Conversely, McKenna [42] also analyzed onshore wind LE in Europe, although many more constraints were considered; including terrain slope, proximity to urban areas, protected regions, and road networks. Despite both investigating onshore wind in Europe, LE results between these two analyses differ to a large degree. Lopez [43] investigated a portfolio of technologies in the United States, including open field PV, CSP and onshore wind. Multiple classifications of protected areas were considered as exclusion constraints for all technologies, including wildlife lands, scenic areas, wildernesses and critical environments. Also, urban areas, wetlands, water bodies and terrain slopes were excluded as well. For onshore wind specifically, Lopez also excluded airports and several definitions of land ownership and, additionally, included a large buffer around most excluded features. For these examples, the LE result is used directly to find the total installable capacity of the region being investigated. This is accomplished by making an assumption about the capacity density of RES technologies installed on the remaining land and simply multiplying by the area found. On top of these, the total producible energy is then estimated by assuming a capacity factor of the resulting total installable capacity.

In MCDM analyses, LE is employed to fully exclude locations within a region, while the MCDM's criterion weighting scheme suggests the likely installation locations within the available area. The procedure used by Watson [35] follows this setup to identify the optimal sites for wind and PV systems in southern England. In this case, LE is defined for both technologies by excluding certain types of agricultural land, 1 km from historically important areas, 1 km from protected landscapes and wildlife areas, 500 m from residential areas and slopes above 10°. For PV, slopes with an aspect not facing between southwest and southeast are also excluded. Following this, resource availability (mean wind speed for wind turbines and solar radiation for PV), as well as the distance from historically important areas, residential areas, wildlife areas, roadways and power lines are all included in an AHP analysis. Ultimately, high suitability locations for the two technologies were identified which showed agreement with the placement of actual systems. The LE plus MCDM pathway has been a common enough theme in the literature that it has been a central issue discussed in multiple literature reviews [4, 15, 16].

LE analyses have also been employed to construct inputs for energy system design models, as showcased by Welder [44] and Robinius [19] in the context of Germany, and by Samsatli [20] in the UK. In each of these cases, an LE analysis determines the maximal capacity of wind turbines in each of multiple sub-regions within the respective study area, followed by an optimization model that determines the capacity distribution that satisfies total energy demand at an overall minimal cost. For the example of Samsatli, 10 exclusion constraints are applied in the LE evaluation, including the exclusion of all locations with a mean annual wind speed below 5 m/s, a slope above 15%, or are designated as a protected environment. Additionally, all locations that are more than 500 m from a road, within 200 m of and over 1.5 km from a power line, within 500 m from residential areas, within 200 m of a river, within 250 m of a forest, within 5 km of an airport and within 5 rotor diameters of preexisting wind turbines are also excluded.

### 2.2. Current State of Land Eligibility Analyses

It is clear that LE analyses are, and will remain, a crucial component of energy-related research. However, despite being of high interest to the research community, the cumulative sum of knowledge regarding LE is ill-structured to meet the demands of spatially-broad contexts in future analyses. Consistency between studies, across a number of dimensions, is the major cause of this situation. This can be readily seen from the few examples provided in the previous section.

Even when investigating the same technology, no two groups of researchers used the same set of exclusion constraints in their LE analysis. Compounding this issue, a closer inspection would reveal not only that completely different datasets are used, but that even the way in which seemingly similar constraints are defined differ from study to study. The studies performed by Samsatli [20] and Watson [35] offer an example of this in the way proximity to settlement areas are handled. Both studies investigate LE for onshore wind turbines in southern England, and furthermore both studies exclude all locations within a buffer distance of 500 m of residential areas. Samsatli, however, specifically identifies these areas according to those defined as *developed land*, while Watson uses the indication of *all dwellings and single properties*. Despite being included for similar reasons (to prevent turbines being place too near to settlements), these two definitions of what constitutes a residential area are distinct from one another and, as a result, their conclusions will differ. This is not to say that either investigation is incorrect, but rather it merely exemplifies an issue that is rampant in LE investigations reported in the literature; inconsistency between individual studies prevents the formulation of a general understanding.

Three sources of inconsistency are specifically identified, namely: the inconsistent use of datasets, inconsistent methodologies and inconsistent criteria definitions. Resch [14] has previously discussed in detail how inconsistent data practices impact the general use of GIS in RES-related studies. This includes the problems associated with data availability, proprietary formats, singular integration methods and an overall lack of standardization. Most notably, Resch points out that investigators must generally perform their own data acquisition and mapping of the relevant information. It is clear to see how, when each group accomplishes this work for their own specific purpose, the datasets used differ in terms of both the source and versions used, as well as in how the datasets are handled. In such a situation, the outcomes of independent studies become increasingly incomparable as the complexity of the LE approaches grow. Although this issue will likely always be a concern to some extent, Resch suggests several avenues that could ease this issue, including the development of unified and generic data models, extended support for open data sources and fitness evaluations of volunteered geographic information (VGI) datasets.

Inconsistency also arises from the implementations themselves. The precise LE methodology utilized by researchers in their analyses are not commonly detailed and, in many cases, are entirely absent from their discussion. In the latter case, a consistent evaluation between LE studies is clearly not achievable. Moreover, the majority of LE studies are conducted with the aid of general GIS applications such as ArcGIS and QGIS, suggesting that the researchers conducting these studies may not themselves be fully aware of the precise operations being utilized. This is an issue, however, as geospatial manipulations often entail situations in which calculation errors are unavoidable, yet a careful choice of where and when to apply geospatial operations can minimize these expected errors. For this reason the chosen chain of operations is a crucial consideration (for which there is no one-size-fits-all solution), which a general GIS application might not account for. Without detailed knowledge of how and when certain geospatial operations are performed, it becomes challenging to reproduce the results of another study or to validate one's own approach.

Finally, criteria definitions and the constraints constructed from these represent the third major source of inconsistency. There are some regions in which exclusion criteria for a particular RES technology have been officially defined, such as that of wind turbines in Austria (used by Hölitinger [1]) and for certain criteria in Greece (used by Latinopoulos [29]). Unfortunately, in general there are no mandates on this topic. Therefore, like the issues faced with the datasets, most researchers must collect this information for themselves prior to conducting their LE analysis. Commonly exclusions from other literature sources are employed, however in some cases the researchers conduct stakeholder questionnaires to identify considerations specifically relevant to their region of study [30, 45, 46]. Nevertheless, in all cases the researchers must at some point rely on their own judgment regarding how to define their constraints. As a result the situation observed previously concerning the differing settlement area constraints used by Samsatli and Watson is very common. Once again, it is clear that the infinite variety by which criteria can be defined, on top of previously mentioned inconsistencies quickly leads to irreconcilable differences when attempting to compare or combine the LE results of various studies.

In rare cases a broad area is evaluated in a single investigation, such as the analyses by the EEA [41], McKenna [18] and Lopez [43] mentioned previously. Unfortunately, the results obtained are still most likely not applicable for use in evaluating energy system development pathways, given that these large scale analyses often do not reflect local preferences and considerations, may not be applicable across different technologies, or at the very least become less relevant over time due to preference shifts and technological advancements.

## 3. LAND ELIGIBILITY FRAMEWORK

Despite the challenges faced by the community, a consistent treatment of LE is nevertheless required for an effective evaluation of energy system developmental pathways. Therefore, the following sections discuss our efforts to relieve some of the previously discussed inconsistencies and, in turn, to simplify and standardize the approach to LE such that the focus can instead be placed on higher order issues. This is accomplished by presenting a general framework for conducting LE analyses which is broken up into three parts. First, a set of criteria definitions are found which are general enough to apply to any land-based RES technology. While these criteria may not cover all considerations for all regions of the world, they nevertheless address the most common and, more importantly, the most impactful criteria as seen in the literature. In this way, even when using differing datasets and investigating different areas, the fundamental considerations made by future LE researchers using this framework can remain constant. Thereafter, a methodological approach to LE is implemented in the Python programming language and has been made available as open source software [47]. The methodology is capable of operating in any geographical context and can accommodate the most common geospatial data formats. With this model we aim to promote the standardization of future LE applications regardless of how criteria are defined and which datasets are used. Finally, a number of existing datasets are standardized in the European context for use in LE analyses. Although limited to Europe, the use of these datasets greatly decreases the required volume of data, computation time, and effort required when conducting an LE analysis and, ultimately, the procedure could be repeated for other geographic areas as well.

As mentioned above, the outlined framework is made available as open source software, including both the model realizing the described methodology as well as the standardized datasets. This model can be found on GitHub under the project name *Geospatial Land Availability for Energy Systems* (GLAES) [47], where version 1.0 corresponds to the version of the code released at the time of this writing. The standardized datasets do not, by default, come with the GLAES model; however, instructions on how to obtain and install them can also be found on the same GitHub page or by contacting the main author of this work.

### 3.1. Criteria Identification

The approach taken to create a generalized criteria set makes the assumption that the most important criteria to consider are already discussed in the collective LE literature; albeit in a myriad of different expressions. Therefore, a sampling of the current LE literature was reviewed. In total, 50 publications [1, 11, 14, 15, 18, 20-24, 26-41, 43, 45, 46, 48-75], representing 55 independent LE analyses, were considered, and the criteria included in each case was tabulated. These studies cover many different technologies, although most investigate either onshore wind or solar thermal plants, while PV and biomass are also present to a lesser degree.

While reviewing the literature sources, four overarching groups were identified that describe the underlying motivation for considering a criterion. These were *physical*, *sociopolitical*, *conservation* and *economic* criteria. Physical criteria are based on the physical characteristics of the land. Among other criteria, this includes terrain slope, the presence of (or, more generally, the distance from) a body of water and the predominant vegetation. Sociopolitical criteria are those derived from the preferences of the local population, generally in response to visual and audible disruptions or safety concerns. This includes distances from settlement areas, roadways, and power lines. Conservation criteria are generally considered to protect the flora, fauna, habitats and facades of designated areas. These are usually determined by national and international organizations and include the exclusion of protected habitats, landscapes, parks, natural monuments, etc. Lastly, economic criteria are considered to identify the locations that are particularly profitable for placing a particular RES technology based on economic or business case considerations. This includes criteria defined from the availability of the primary resource, the costs associated with connecting the technology to the power grid, as well as other costs associated with making the location accessible for construction and maintenance (i.e., by building a road).

All observed criteria were then categorized into their respective motivation groups. As mentioned previously, the literal expression of criteria within each study are commonly unique; nevertheless, the criteria were generalized as much as possible. As an example of this, if one study chose to consider "distance from settlement areas" as a criterion, such as that used by Höltinger [1], while another chose to use "distance from urban areas and traditional settlements," such as that employed by Latinopoulos [29], both instances would be counted as instances of a "settlement area proximity" criterion. In this way the consideration rate of each criterion is recorded. During this phase, the nature of each criterion is determined as well. Two general types of criterion are identified: a *Value* type simply indicates a direct property of the location in question (like elevation or land cover) while a *Proximity* type indicates a location's distance from some feature (such as a road or a settlement). Additionally, when appropriate, the implied desirability for each criterion is found where *High* implies that higher values are more desirable while *Low* implies the opposite, and *Range* indicates that only a specific range of values are typically considered desirable. Finally, as these criteria are used to define both the exclusion constraints for LE analyses as well as the decision factors in MCDM analyses (which many of the reviewed studies also performed), a final outcome of the review is each criterion's consideration rate when specifically considered as an exclusion constraint.

### 3.1.1. Criteria Description

In the end, 28 general criteria were identified and are shown in Table 1. The following sections describe each of the indicated criteria in more detail and provide some common arguments for their consideration in the review literature. A table showing the criteria which were considered by each of the reviewed reports can be found in Appendix A.

#### Sociopolitical

Within the sociopolitical motivation group, 13 general criteria were identified. The most commonly considered criteria in this group was *distance from settlement areas* where, as indicated in Table 1, distances closer to settlements are considered less desirable. This criterion is generally used to account for both safety issues related to RES technology, as well as to account for visual and audible disruptions to the local population (such as glare from PV panels and noise from wind turbines). Although many studies utilized distance from settlements in a general sense, many also distinguished several sub-definitions; most commonly, this included the distance from urban (or otherwise densely populated) settlements, as well as from rural (sparsely populated) settlements. *Distance from*

*airports* was the next most commonly included criterion, which was generally considered as a safeguard against dangers to airline passengers (such as increased turbulence in wind turbine wake, or the glare of PV panels affecting pilots), although could also account for effects of airplane turbulence on the RES technologies (such as increased soiling on PV panels). Two sub-criteria were also found for this criterion, namely the distance from large and commercial airports and from smaller airfields. The *distance from roadways* follows, accounting for safety concerns related to placing an RES technology too close to a roadway (such as potential ice throws and structural failures of wind turbines or, again, glare from PV panels or CSP reflectors). Once again, sub-criteria are found here: the distance from primary roads (such as highways and federal roadways) and from secondary roadways (such as those connecting rural settlements). The *distance from agricultural areas* refers to distances from agriculture areas such as pastures and farmland, which are considered to prevent the installed technology from interfering with food supply. This criterion was generally only considered for PV, CSP and biomass systems, as these technologies either interfere or compete with the resources needed by crops, while wind turbines were commonly claimed to be compatible with most types of agriculture [21]. Similar to roadways, the *distance from railways* was considered a safeguard against dangers to passengers and damage to infrastructure and was most commonly considered for wind turbines. Likewise, *distance from power lines* was also commonly included to avoid damage to infrastructure and was, once again, most commonly considered for wind turbines. *Distance from historical sites* was typically included to prevent installations too close to locations of cultural and historic importance, so as to avoid disruptions to visitors to such sites and to prevent the devaluing of a site's significance. This included battlegrounds, castles, monuments, religious sites, and others. No sub-criteria are defined here, as a consensus could not be found in regard to what constitutes a historically significant site. *Distance from recreational areas* was also generally considered to avoid disruption and danger to visitors. Three sub-criteria were identified here, namely: distance from campgrounds, from urban parks (or 'green' parks) and from tourist attractions. *Distance from industrial areas* was considered to prevent the placement of a technology too close to industrial areas in order to protect equipment and workers. *Distance from mining sites* was included to prevent placement too close to mining sites and to protect equipment and workers, and also given that the ground may be unstable in these areas. The general implication in this case saw siting placements further away from mining sites as preferable, although some studies actually used the opposite logic with regard to PV plants, stating that the areas around mining sites presented ideal locations for PV [21]. The *distance from radio towers* was included to prevent damage to infrastructure and avoid disruption of the signal broadcast from the tower; in general, this constraint was only considered for wind turbines. Similar to distance from power lines, *distance from gas lines* was considered to avoid damage to infrastructure. As the last criterion in this group, *distance from power plants* was generally included in order to protect equipment and workers in the area.

*Table 1: Typical criteria employed for LE analyses as found in the literature*

| Criterion | Inclusion Rate (%) | | Type | Preference |
|---|---|---|---|---|
| | *General* | *Constraint* | | |
| **Sociopolitical** | | | | |
| Settlements | 85 | 84 | Proximity | High |
| Airports | 55 | 51 | Proximity | High |
| Roadways | 53 | 51 | Proximity | High |
| Agricultural Areas | 44 | 29 | Proximity | High |
| Railways | 33 | 31 | Proximity | High |
| Power Lines | 31 | 27 | Proximity | High |
| Historical Sites | 27 | 25 | Proximity | High |
| Recreational Areas | 20 | 18 | Proximity | High |
| Industrial Areas | 18 | 18 | Proximity | High |
| Mining Sites | 15 | 11 | Proximity | High |
| Radio Towers | 9 | 7 | Proximity | High |
| Gas Lines | 7 | 5 | Proximity | High |
| Power Plants | 4 | 4 | Proximity | High |
| **Physical** | | | | |
| Slope | 69 | 65 | Value | Low |
| Water Bodies | 64 | 64 | Proximity | High |
| Woodlands | 40 | 33 | Proximity | High |
| Wetlands | 31 | 27 | Proximity | High |
| Elevation | 18 | 15 | Value | Low |
| Land Instability | 16 | 15 | Proximity | High |
| Ground Composition | 15 | 7 | Value | - |
| Aspect | 7 | 5 | Value | Range |
| Vegetation | 15 | 0 | Value | - |
| **Conservation** | | | | |
| Protected FFH | 82 | 75 | Proximity | High |
| Protected Areas | 67 | 65 | Proximity | High |
| **Economical** | | | | |
| Resource | 64 | 38 | Value | High |
| Access | 45 | 25 | Proximity | Low |
| Connection | 47 | 24 | Proximity | Low |
| Land Value | 13 | 5 | Value | - |

<u>Physical</u>

The physical motivation group is characterized by nine general criteria. Most commonly, the *terrain slope* was considered to measure the average slope of the terrain throughout the area relevant to each location. This criterion was applied to all RES technologies, although the terrain slope in the north-south direction, a sub-criterion of the terrain slope, was only applied to solar and biomass technologies. Following this, *distance from water bodies* was included on several grounds, including: to protect the installation site from water damage during rainy periods, to avoid contamination of water streams during the construction process and to prevent disturbance of the wildlife dependent on these water bodies. The treatment of water bodies varied heavily between sources however, so three sub-criteria have been extracted from this group. The distance from stagnant water bodies indicates a location's distance from lakes, reservoirs and any other bodies characterized by still water. The distance from running water bodies indicates a location's distance from rivers, streams, canals and other bodies characterized by flowing water. Finally, distance from coasts refers to a location's distance from the nearest coastline, which is treated separately from stagnant water bodies, as alternative issues must be considered (such as the tide and corrosion from sea spray). *Distance from woodlands* indicates a location's distance from the nearest forest and was considered for all technologies, as the presence of nearby woodlands can affect resource availability (such as slowing wind speeds and blocking solar irradiance) and the systems themselves can adversely impact local wildlife. *Distance from wetlands* was considered to avoid terrain unsuitable for construction and ensure the avoidance of riparian zones and wetland areas. Mean *Elevation* was included under the reasoning that suitability for RES technologies decreases at extreme elevations due to inaccessibility, instillation costs and diminishing resources (in the form of lower air density and increased cloud coverage, for example). The *distance from land instability* criterion was considered to prevent installation too close to areas which are prone to landslides, mud slides, or earthquakes in order to avoid the damage these events would inflict on an installation site. *Distance from ground composition* was considered as a criterion to account for certain soil and ground compositions that can affect an RES installation. A notable sub-criterion is given by distance from sand coverage (such as beaches and sand dunes), as the sand might not provide stable land to build on and, when carried by the wind, can cause erosion, soiling (in the case of solar technologies) and disturb plant growth (in the case of biomass). The *aspect* criterion was considered to account for how the direction of the terrain slope affects an RES installation; for example, how slopes facing in a northerly direction will experience a diminished solar resource, or how slopes facing away from the predominant wind direction will be subject to wind shading. Finally, the *vegetation* criterion was commonly considered to account for how local flora interact with an RES installation by means of, for example, damage caused by root growth, as the RES installation could disturb the ecosystem of particularly vulnerable vegetation via reduced sunlight (such as by PV panel coverage) and by competition for nutrients (in the case of biomass).

*Conservation*

The conservation motivation group is characterized by only two main criteria; however, multiple sub-criteria were found in each case. *Distance from protected flora, fauna and habitats (FFH)* was considered to prevent RES installations from adversely affecting vulnerable ecosystems and commonly relied on national or international designations. As sub-criteria, this criterion is separated into distance from designated habitats (which include bird and bat areas), from animal migration routes, from biospheres and from wildlife refuges. The *distance from protected areas* criterion was included to prevent RES installations from interfering with protected areas outside of ecological concerns. Sub-criteria in this group include distance from designated landscapes, from protected designated parks, from designated nature reserves and from natural monuments.

*Economical*

Within the economical motivation group, four general criteria were identified. The *resource* criterion served as an indicator for how much energy an RES installation at a location would potentially produce. Two sub-criteria were identified according to the quality that was most relevant to the RES technology being investigated: annual mean wind speed in the case of wind turbines and average daily global horizontal irradiance for solar and biomass technologies. The *access* criterion was commonly included to account for costs associated with providing access to the RES installations during both the construction phase and for maintenance. The implication being that locations that are less accessible with existing infrastructure would incur higher costs as new roadways must be constructed. More often than not, the access criterion was defined according to distance from the nearest roadway, although considering that the opposite implication is used compared to the roadway consideration in the sociopolitical motivation group (where desirability increases with distance) and that the two operate at completely different scales (many kilometers versus hundreds of meters), the access criterion is treated separately than road proximity. Similarly, the *connection* criterion was considered to account for the installation costs associated with connecting the RES installation to the energy network, the implication being that desirability decreases as distance increases as new transmission infrastructure must be built spanning larger distances. When wind turbines and solar technologies were considered, this criterion was defined by the distance to the nearest electricity grid line, while when biomass was considered the distance to the nearest natural gas pipeline was used. Once again, the connection criterion is treated separately from the power lines and gas network criteria in the sociopolitical motivation group because of the opposite implication and different operating scales. Finally, the *land value* criterion was used to account for the costs associated with obtaining construction rights at each location, as estimated by the current owners of the land. For similar reasons, as discussed with respect to the distance from historical sites, when this criterion was considered, the definitions of land ownership were highly specific to the local region being investigated and, as such, no general consensus could be extracted.

3.1.2. Constraint Ranges

In total 28, general criteria were identified as being utilized somewhere within the work flow of the reviewed studies. However, as each study also included an LE investigation, Table 1 also provides the inclusion rate of each criterion specifically within the study's LE analysis. As was previously discussed, when criteria are expressed as an exclusion constraint, a value threshold, range, or subset is used to differentiate between which locations are eligible and which should be excluded. In the vast majority of cases, a threshold is given, and therefore in order to get an idea of what threshold values are typically used for each constraint, these values were recorded when reviewing the literature and are summarized in Table 2. For each of the indicated criteria (and in many cases sub-criteria when a consensus within the literature could be identified), this table gives a typical threshold value that resulted in low exclusions, a threshold that resulted in high exclusions and a typical threshold value used across all observations. Although the low and high exclusion values contain the vast majority of all applications for each constraint, the typical exclusion reported in this table is not intended for any single particular technology, as the chosen threshold for an analysis depends heavily on the technology chosen, the intended application (such as grid-connected or off-grid), and the preferences of the region being investigated. Instead, the typical value is given to indicate the scale at which these constraints require their underlying criterion to offer detailed information; which will come into play in the following discussion, where standardized datasets are produced for Europe.

**Table 2: Typical exclusion constraint expressions** used for LE analyses as seen in the literature

| Group<br>Criterion<br>*Sub-criterion* | Excludes | Low | Typical | High | Unit |
|---|---|---|---|---|---|
| **Sociopolitical** | | | | | |
| Settlements | distances below | 0 | 800 | 2,000 | m |
| *Urban Settlements* | distances below | 0 | 1,000 | 3,000 | m |
| *Rural Settlements* | distances below | 240 | 500 | 2,000 | m |
| Airports | distances below | 0 | 5,000 | 8,000 | m |
| *Large & Commercial* | distances below | 0 | 5,000 | 25,000 | m |
| *Airfields* | distances below | 0 | 3,000 | 8,000 | m |
| Roadways | distances below | 50 | 150 | 500 | m |
| *Primary* | distances below | 50 | 200 | 500 | m |
| *Secondary* | distances below | 50 | 100 | 500 | m |
| Agricultural Areas | distances below | 0 | 50 | 240 | m |
| Railways | distances below | 50 | 150 | 500 | m |
| Power Lines | distances below | 100 | 200 | 240 | m |
| Historical Sites | distances below | 500 | 1,000 | 3,000 | m |
| Recreational Areas | distances below | 0 | 250 | 500 | m |
| Leisure & Camping | distances below | 0 | 1,000 | 3,000 | m |
| Tourism | distances below | 500 | 800 | 1,000 | m |
| Industrial Areas | distances below | 0 | 300 | 500 | m |
| Mining Sites | distances below | 0 | 100 | 500 | m |
| Radio Towers | distances below | 400 | 500 | 600 | m |
| Gas Lines | distances below | 100 | 150 | 300 | m |
| Power Plants | distances below | 100 | 150 | 200 | m |
| **Physical** | | | | | |
| Slope | values above | 30 | 10 | 1 | ° |
| Water Bodies | distances below | 0 | 300 | 3,000 | m |
| *Lakes* | distances below | 100 | 400 | 4,000 | m |
| *Rivers* | distances below | 0 | 200 | 400 | m |
| *Coast* | distances below | 0 | 1,000 | 3,000 | m |
| Woodlands | distances below | 0 | 300 | 1,000 | m |
| Wetlands | distances below | 0 | 1,000 | 3,000 | m |
| Elevation | values above | 2,000 | 1,800 | 1,500 | m |
| Land Instability | distances below | 0 | 200 | 500 | m |
| Ground Composition | | | | | |
| *Sand* | distances below | 0 | 1,000 | 4,000 | m |
| **Conservation** | | | | | |
| Protected FFH | distances below | 0 | 500 | 2,000 | m |
| *Habitats* | distances below | 0 | 1,500 | 5,000 | m |
| *Biospheres* | distances below | 0 | 300 | 2,000 | m |
| *Wildernesses* | distances below | 0 | 1,000 | 4,000 | m |
| Protected Areas | distances below | 0 | 1,000 | 3,000 | m |
| *Landscapes & Reserves* | distances below | 0 | 500 | 3,000 | m |
| *Parks & Monuments* | distances below | 0 | 1,000 | 3,000 | m |
| **Economical** | | | | | |
| Resource | | | | | |
| *Wind speed* | values below | 4.0 | 4.5 | 7.0 | m/s |
| *Irradiance* | values below | 4.5 | 5.0 | 5.5 | $\frac{kWh}{m^2 day}$ |
| Access | distances above | 45,000 | 5,000 | 1,000 | m |
| Connection | distances above | 40,000 | 10,000 | 1,000 | m |

### 3.2. Land Eligibility Implementation

The desired LE model approach is one that, when provided with suitable datasets, can realize exclusion constraints (defined however the LE investigators see fit) and combine these exclusions into a unified LE result. As discussed in Section 2.2, this should be accomplished in a manner that is scalable to large geographical areas, minimizes expected sources of error and is methodologically transparent. A model has been designed that satisfies these aims and, as previously mentioned, is henceforth offered as open source software hosted on GitHub under the project named Geospatial Land Eligibility for Energy Systems (GLAES) [47] . The model has been implemented in the Python 3.4 programming language, with primary dependencies on the Geospatial Data Abstraction Library (GDAL) [76] for geospatial operations and on the SciPy [77] ecosystem for general numerical and matrix computations, both of which are also open source projects.

#### 3.2.1. Raster vs. Vector

Before discussing the underlying LE methodology, however, a simplified introduction to two common formats of geospatial data is provided as they become relevant in the following discussions. The first of these, a *raster* dataset, communicates geospatial data in the form of an image. Rasters are defined in the context of a spatial reference system (SRS) (which maps an X and Y coordinate to a specific location on the Earth's surface), a rectangular boundary (extent) described by that SRS, and a pixel resolution (which determines to how much land is associated with each pixel). The result of this definition describes a rectangular grid of pixels, each of which contains a specific value representing some property of its associated location. A straightforward case of a raster dataset is given by elevation, where each pixel in the raster dataset would give the mean elevation of each pixelated location in the raster's extent. In a more complex setup, a raster's pixel values can serve as indicators for a more intricate property. Land cover datasets, for example, will routinely use integer values to indicate certain classes of land cover. A pixel value of "1" could indicate that the location is predominantly covered by, for instance, woodlands, a value of "2" could indicate grasslands, "3" could indicate urban coverage, and so forth. Rasters are most often applied to quantities that are definable for all conceivable locations which, in addition to elevation and land cover, can include population density, mean temperature, and average wind speed.

Along with raster datasets, a *vector* dataset constitutes the other common method of communicating geospatial information. Vector datasets are also defined according to a particular SRS, but instead of representing their information as a continuous grid of values they instead provide a collection of explicitly defined features and associated attributes. Each feature in a vector dataset is given by a geometry which, dependent on the type of feature being represented, can either be a point, a line, or a polygon. A point feature could indicate, for example, specific landmarks, while line features would be used to indicate pathways, such as roads and rivers. Polygon features are used to indicate an area, such as the boundaries of an administrative area or of a river's catchment zone. Regardless of how a feature is geometrically represented, all features are also associated with a list of attributes that provide more detailed information about each feature. Using a vector dataset describing roadways as an example, the features may each have an attribute for the road's name, its speed limit, whether or not it is a one-way street and how many lanes it has. Attributes can be given as numerical values or character strings, and all features within a dataset will share the same number of attributes.

#### 3.2.2. Region Definition

The first step within the constructed model involves creating a regional context over which the future computations will operate. Not only does this context determine the geographical area that is extracted from the various exclusion-datasets, but it also defines the raster characteristics (resolution, spatial reference system and extent) of the finalized LE result. Defining this context requires at least three parameters: the boundary of the region to be investigated, the desired output SRS and the desired pixel resolution. There is no methodologically-inherent limitation on the geographic areas, SRSs and resolutions that can be represented by the described model; however, in the case of LE analyses the output SRS would usually be an equal-area projection which preserves relational distances and the resolution should be small enough to capture local details (later analyses in this report use 100 m).

When given the necessary inputs, the model represents the regional context as a region mask (RM), which is exemplified in Figure 1. The model first transforms the regional geometry to the desired output SRS and records the enveloping extent (the smallest rectangular extents that contain the region while still fitting the given resolution, as shown in the figure by *xMin, yMin, xMax,* and *yMax*). With the output extent, resolution and SRS in hand, a raster dataset comprised of a single band of boolean values is created with these characteristics. The pixels of this raster which are mostly within the original region definition are given the value 1, while all other pixels are given the value 0. In essence, the RM serves as a basis onto which all exclusion information will be translated, at which point the RM's boolean mask is used to easily determine which of those pixels are within the regional area.

After creating the RM, an output *availability* matrix is also initialized. This matrix shares the same characteristics (SRS, extent, resolution and datatype). All pixels in the availability matrix are initially filled with the value 1 indicating that, before any exclusion has been applied, all locations are considered to be 100% available. Henceforth, the various exclusion datasets will be processed and the indicated exclusions removed from the availability matrix via element-wise logical operations.

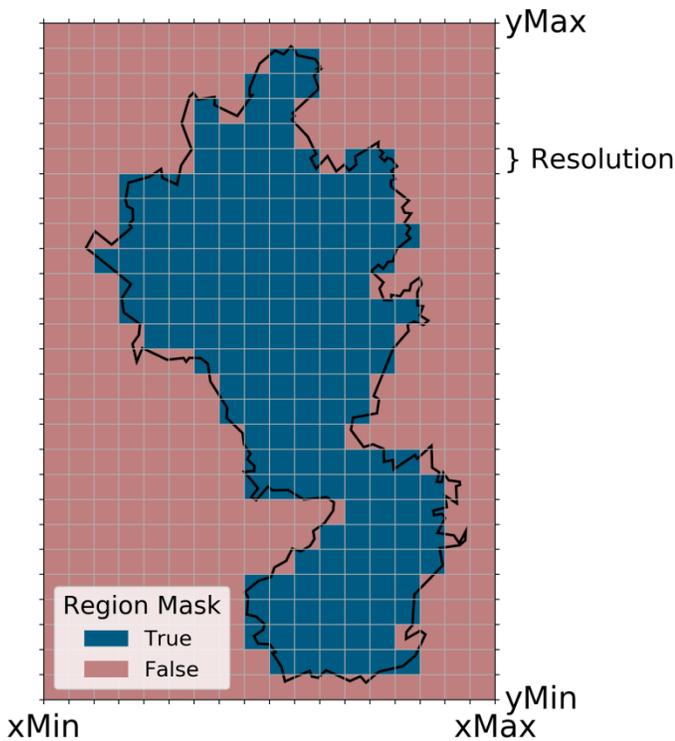

**Figure 1: Region Mask** *defined from a region definition, SRS and resolution.*

### 3.2.3. Exclusion Indication

All applications of the exclusion procedure generate a 2-dimensional boolean matrix, matching the characteristics of the RM. In these matrices, true-valued pixels indicate that the associated locations should be excluded from the final availability matrix. Figure 2 provides a simplified flowchart of the implemented procedure. In general, the exclusion procedure functions as follows: first, an indication of the locations to be excluded is made in the dataset's original definition (or as close to it as possible), then those indicated locations are translated into the RM's characteristics, after which the indicated pixels are excluded from the availability matrix. However, considering that raster and vector source represent spatial information in drastically different ways, the procedure must follow one of two initial tracks, depending on the type of input dataset. These two initial tracks are described below.

For raster sources, the only parameter required is the indication value or range of indication values that should be marked for exclusion, although an optional buffer distance parameter can be provided as well. The first objective is to generate a new boolean-valued raster source, indicating the to-be-excluded pixels in the source's original SRS and resolution. To minimize memory usage, however, the original source's extent is always clipped to the smallest extent (defined in the source's SRS), which contains the RM's extent. Indication is accomplished by considering each pixel in the original source and determining if its value is equal to, or is otherwise within the range of, the given indication value parameter. This new source, which now contains boolean values indicating whether or not each pixel should be excluded, is then warped using the bilinear method to a new raster in the RM's extent and SRS. By performing the warping procedure after the indication step, the unavoidable error introduced from warping is minimized compared to the reverse order.

When a vector dataset is provided, there are no required inputs; however, an optional feature filter, buffer distance, and buffer method parameters can be provided. The most appropriate parameterization to use depends greatly on the characteristics of the dataset that is being processed. As a first step, the original source is spatially filtered and a new vector source is generated that contains only the features of the original vector which overlap the RM's extent. If a feature filter has been provided, another new vector source is produced that only contains the extent-filtered vector's features that also pass the feature filter. Following this, if a buffer distance parameter is provided and the buffer method parameter indicates that the geometry-based method should be used (which is the default pathway in this case), yet another intermediate vector source is then produced. First, the filtered source's features are segmentized to a distance equivalent to half of the RM's resolution, transformed to the output SRS and then buffered by the given buffer distance. By segmentizing the original geometries, the error introduced when transforming to the output SRS is minimized, as the new geometry will follow the contours of the original more closely than if the geometry was transformed directly. Moreover, since the buffer distance is provided in units of the output SRS, applying the buffer distance to geometries that are expressed in the output SRS ensures that an identical buffer distance is applied everywhere. In any case, the latest intermediate vector source is comprised of geometries that indicate the areas to be excluded. This is then rasterized onto an intermediate boolean-valued raster in the RM's SRS and extent in the same manner described when generating the region mask.

Regardless of the original source's type, at this point (indicated with a star in the center of Figure 2) an intermediate boolean-valued raster has been produced in the RM's characteristics. If a buffer distance parameter was provided and the original source was of the raster type or if the area-based method was indicated in the vector case, then a buffer distance must be applied around all of these indicated pixels. To accomplish this, the intermediate raster is polygonized, converting all contiguous true-valued regions into geometries described in the output SRS. These geometries are then buffered by the given buffer distance, followed by another rasterization of these geometries into a new boolean-valued raster matching the RM's characteristics. Finally, application onto the availability matrix is performed via an element-wise 'or' operation between an inverted version of this output raster and the availability matrix.

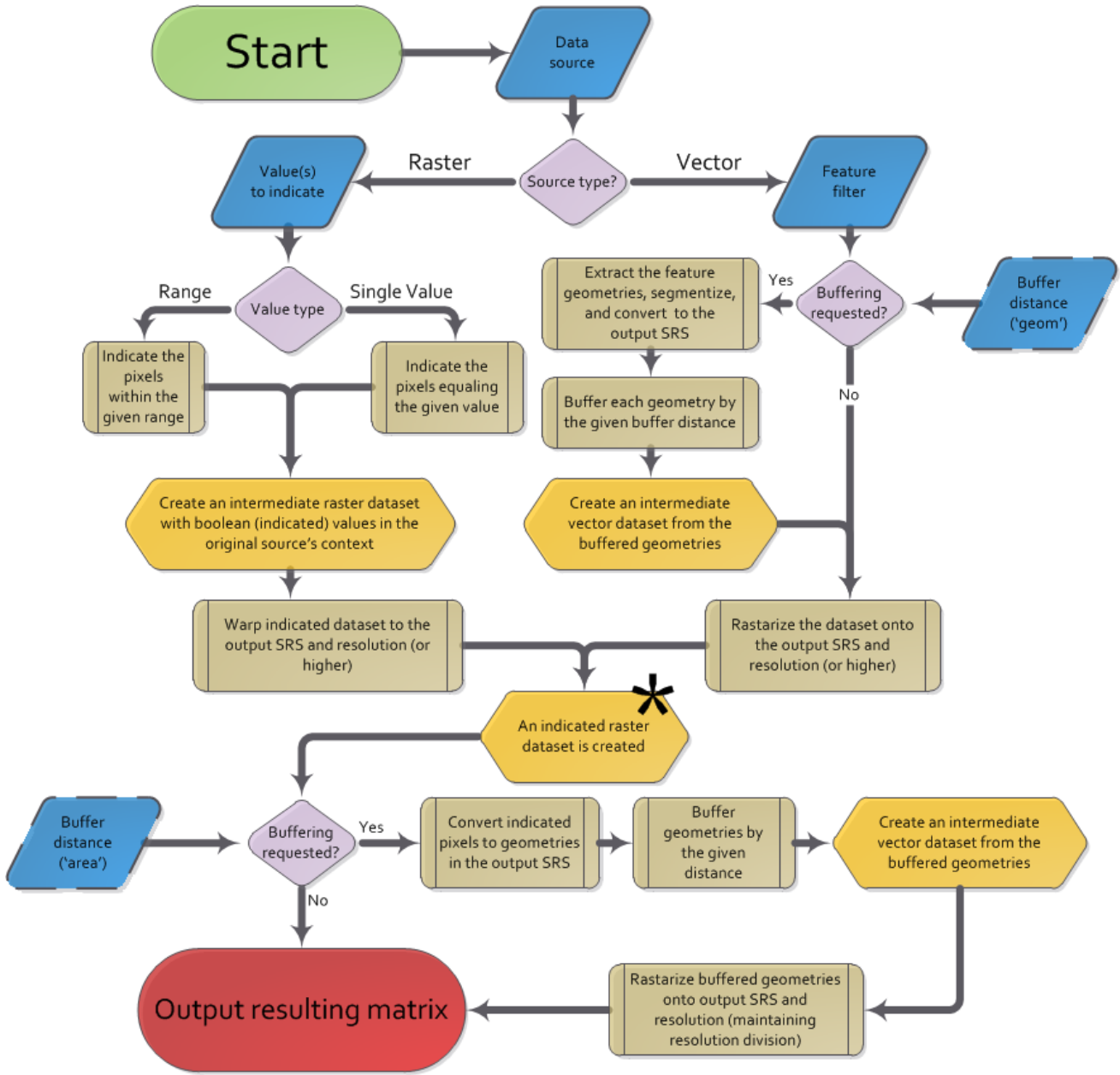

***Figure 2: Exclusion flow chart*** *showing the procedural for indicating the exclusion pixels from a given dataset according to the user's preferences.*

### 3.2.4. Application

With consecutive applications of this procedure using a multitude of exclusion datasets, it can be seen how the availability matrix is updated with each iteration; which always results in a smaller amount of available pixels. Furthermore, the described procedure is general enough that, with proper parameterization, it can be applied to any raster or vector dataset[1] and to any geographical area. Ultimately, the availability matrix will be filled with boolean values, where a value of 1 implies that the pixel remains available after all exclusion indications and a value of 0 implies the pixel has been excluded by at least one exclusion constraint.

### 3.3. Dataset Standardization

Resolving the issue of inconsistent underlying datasets, as discussed in detail by Resch [14] in the general sense would require a collective effort far beyond the capabilities of the authors of this work. Nevertheless, with the advent of broad context VGI and institutionally-funded open data sources, most of the previously defined criteria (Table 1) have become expressible across the European context. For this reason, an effort was made to process a number of these available data sources and, in turn, create a standardized set for the purpose of LE analyses in Europe. The use of these standardized datasets, now referred to as *Priors*, has the added benefit of reducing the overall size of the data required for LE analyses, in addition to drastically reducing the processing time. Furthermore, these datasets are also made freely available along with the model described previously on the GitHub repository [47] in order to promote consistency in the dataset usage for LE studies in Europe.

---

[1] As long as it is one GDAL can manipulate

*Table 3: **Prior datasets** produced over the European context, with small description, total number of edges, and sources used in their creation*

| | Prior Name | Description | Edges | Sources |
|---|---|---|---|---|
| **Sociopolitical** | Settlement proximity | Distance from all settlement areas | 38 | [25] |
| | Settlement urban proximity | Distance from urban settlement areas | 38 | [78] |
| | Airport proximity | Distance from airports | 29 | [25, 79] |
| | Airfield proximity | Distance from airfields | 29 | [25, 79] |
| | Roads proximity | Distance from all roadways | 33 | [80] |
| | Roads main proximity | Distance from major roadways | 37 | [80] |
| | Roads secondary proximity | Distance from secondary roadways | 37 | [80] |
| | Agriculture proximity | Distance from all agricultural zones | 20 | [25] |
| | Agriculture arable proximity | Distance from arable agricultural zones | 20 | [25] |
| | Agriculture permanent crop proximity | Distance from permanent crop areas | 20 | [25] |
| | Agriculture heterogeneous proximity | Distance from mixed-use agricultural areas | 20 | [25] |
| | Agriculture pasture proximity | Distance from pastures and grazing areas | 20 | [25] |
| | Railway proximity | Distance from railways | 34 | [80] |
| | Power line proximity | Distance from power lines and stations | 36 | [80] |
| | Leisure proximity | Distance from leisure areas and public parks | 20 | [80] |
| | Camping proximity | Distance from camp sites | 20 | [80] |
| | Touristic proximity | Distance from well-known touristic spots | 20 | [80] |
| | Industrial proximity | Distance from industrial units | 20 | [25] |
| | Mining proximity | Distance from mining sites | 20 | [25] |
| **Economical** | Slope threshold | Average terrain slope | 61 | [81] |
| | Waterbody proximity | Distance from all water bodies | 20 | [82] |
| | Lake proximity | Distance from lakes | 20 | [65] |
| | River proximity | Distance from probably river routes | 21 | [83] |
| | Ocean proximity | Distance from coast lines | 23 | [25] |
| | Woodland proximity | Distance from all woodlands | 20 | [25] |
| | Woodland deciduous proximity | Distance from deciduous (broad leaf) woodlands | 20 | [25] |
| | Woodland coniferous proximity | Distance from coniferous (needle-leaf) woodlands | 20 | [25] |
| | Woodland mixed proximity | Distance from mixed species woodlands | 20 | [25] |
| | Wetland proximity | Distance from all wetlands | 20 | [25] |
| | Elevation threshold | Average terrain elevation | 41 | [81] |
| | Sand proximity | Distance from predominantly sandy areas | 21 | [25] |
| **Conservation** | Protected habitat proximity | Distance from protected habitats | 20 | [84] |
| | Protected bird proximity | Distance from designated bird areas | 20 | [84] |
| | Protected biosphere proximity | Distance from protected biospheres | 20 | [84] |
| | Protected wilderness proximity | Distance from protected wilderness | 20 | [84] |
| | Protected landscape proximity | Distance from protected landscapes | 20 | [84] |
| | Protected reserve proximity | Distance from protected nature reserves | 20 | [84] |
| | Protected park proximity | Distance from protected parks | 20 | [84] |
| | Protected natural monument proximity | Distance from protected natural monuments | 20 | [84] |
| **Economic** | Windspeed 50m threshold | Average annual wind speed at 50m | 80 | [85] |
| | Windspeed 100m threshold | Average annual wind speed at 100m | 80 | [85] |
| | DNI threshold | Average direct normal irradiance per day | 80 | [86] |
| | GHI threshold | Average global horizontal irradiance per day | 80 | [86] |
| | Access distance | Distance from all roadways | 33 | [80] |
| | Connection distance | Distance from power lines and stations | 36 | [80] |

### 3.3.1. Prior Dataset Description

Each of the Prior datasets represents exactly one criterion, or sub-criterion, shown in Table 1. Furthermore, all Priors take the form of a single-banded, byte-valued raster dataset defined over the European context[2]. They are expressed in the EPSG:3035[3] SRS and possess a resolution of 100 m by 100 m. Byte values were chosen to conserve the overall size of each individual dataset, but restrict the datasets to only containing integer values between 0 and 255. Therefore, instead of representing criteria values directly, each value in the Prior datasets (except 255 and 254) is associated with a given set of strictly increasing criteria thresholds, referred to as *edges*, indicating the minimal edge that includes each pixel. The value associated with each edge is given by its index position in the ordered set of edges, where a value of zero identifies the first edge (resulting in the fewest indicated pixels), while the highest value identifies the pixels indicated by the final edge. In all cases, the value 255 is used as a "no data" value that indicates offshore areas or locations that are otherwise unprocessed. Likewise, a value of 245 means "no indication", meaning the pixel was not indicated by any edge value.

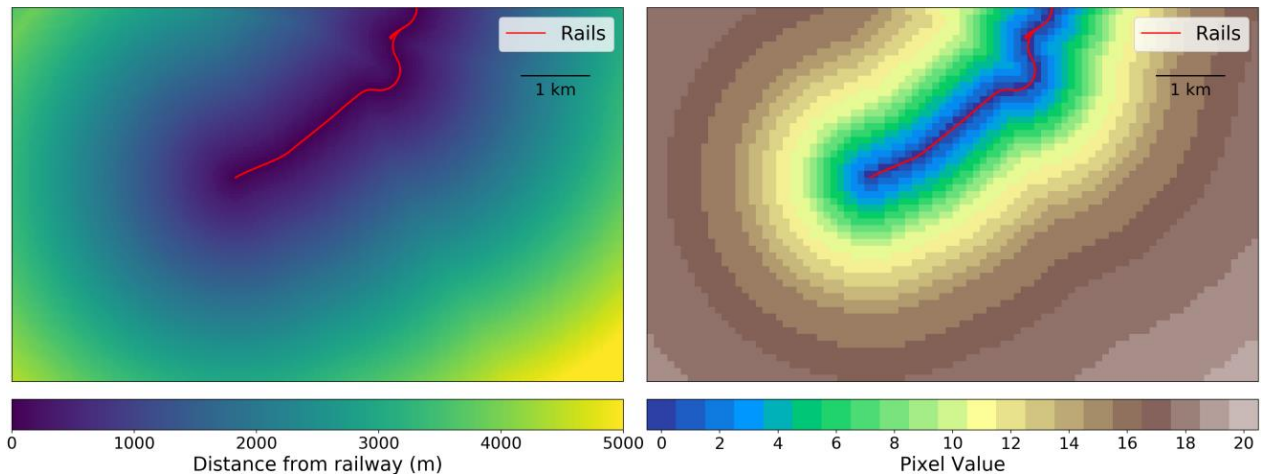

***Figure 3: Prior Example*** *of the distance from the railways criterion*

Although the exact production method of each Prior dataset is unique, the general procedure involves deciding on a set of edges and recording the first indicating edge for each location in the European context. For the criteria and sub-criteria expressible as a Prior, the edge values chosen span the range of exclusion constraint thresholds shown in Table 2 and are relatively more detailed wherever the typical threshold is found. Table 3 provides a list each of the 45 generated Priors and also gives a short description, indicates the number of edges included, and finally identifies the dataset underlying their creation. A detailed description of each of the Priors, including their production method, can be found in Appendix B, while a table of the literal edge values used can be seen in Appendix C.

### 3.3.2. Prior Dataset Usage

Although the Prior datasets do not directly represent the criteria values at each location, but rather contain the index of the first containing edge, the framework described here can translate from an edge index to a Prior's original criterion value. Therefore a user is only required to provide criteria values that must be excluded and the framework will translate these values into appropriate indexes. Ultimately, all Priors behave exactly like the raster datasets described in Section 3.2, where a minimal or maximal threshold value is given to indicate which locations should be excluded. Both a minimal and maximal threshold value can also be provided if an indication range is desired.

Figure 3 gives an example of a Prior dataset, which describes this setup using the *distance from railways* criterion as an example. The figure shows the path of a railway (red line) entering from the top of the scene. The image on the left shows, for each point in the space surrounding the railway, the distance of that location from the nearest point on the red line, while on the right the corresponding pixel values in the railways Prior dataset are shown. The specific edge values shown in this instance are 0, 100, 200, 300, 400, 500, 600, 700, 800, 900, 1000, 1200, 1400, 1600, 1800, 2000, 2500, 3000, 4000 and 5000 meters from the railways. In this sense, an edge of 0 refers to the pixels that are indicated by the unprocessed railway geometries (as in, only indicating the pixels that lie on the railway's path). Considering that the pixels themselves have an inherent resolution of 100 m, however, the Prior dataset cannot exactly represent features smaller than 100 m. All other edges indicate the pixels that are within an area comprised of the railways plus a buffer distance of the given edge. A pixel's value identifies the smallest edge that indicates the given pixel, however these pixels would of course also be indicated by edge values greater than the identified edge; for example, a pixel that is indicated by 500 meters from the railways would also be indicated when considering 600 meters from the railway. It can be seen in the figure that there is much more detail represented by the first 10 edge values, where every 100 meters are indicated, than in the last edge values, where only every 1000 meters is indicated. This setup is chosen to provide adequate detail for typical applications of the distance from the railways criterion while not bloating the resulting dataset's file size.

The granularity of the Prior datasets limit the accuracy of the information they provide when criteria values differ significantly from the edge values used, and therefore they should only be employed when a more detailed dataset is unavailable. Nevertheless, the values they contain can still be used directly for LE analysis, as they contain adequate detail in the ranges typically used for these analyses. The following sections will reveal this by directly validating the use of the Prior datasets by means of recreating LE results reported in the literature. In addition to this, however, more precise criteria values can be estimated by finding the inner boundary of two adjacent edge regions and interpolating between them. While this would not provide an exact recreation of the original dataset, it will be approximate to the scale relevant to LE analyses. This interpolation feature has also been built into the framework and can be used directly.

## 4. VALIDATION

In view of validating the described framework, eight studies that include LE analyses were chosen for replication. To qualify for replication, each study's LE analysis had to have been conducted within Europe and where the majority of exclusion constraints are closely expressible in terms of the Prior datasets. It was decided that the validation effort would focus on the use of the Prior datasets for reasons of both simplicity and to emphasize how these Prior sources can produce acceptable results despite their lower granularity compared to the raw underlying datasets. Seeing as how the Prior datasets are fixed on their criteria definitions and limited to information contained within their original datasets, they will likely not match perfectly to the datasets and criteria definitions used in the various replication studies. Therefore, if the result of combining several Prior datasets to recreate previous LE analyses comes close to the reported values, then the overall framework will be considered valid as a more specific choice of datasets and subtle alterations of criteria definitions will surely improve the end result.

The procedure taken for each replication was roughly the same in all cases. Following along with the methodology described in Section 3.2, a region used by the replication authors is initialized, followed by the application of multiple Prior datasets, along with associated criteria value thresholds. The datasets and thresholds used in each case are given in Table 4. The number of available pixels in the replication region is compared against the total number of pixels in the replication region, yielding the percentage of available land remaining. Table 4 also displays the result of each replication effort.

### 4.1. Validation Studies

For each replicated study, a short introduction is given, after which the significant differences from the Prior datasets are summarized. When a course of action was taken to try to account for these discrepancies, the reasoning for this is given as well.

The LANUV [60] study investigated onshore wind LE in North Rhine Westphalia, Germany. Typical wind-relevant exclusion constraints were considered, including distances from roads, settlement areas, airports, railways, power lines, rivers and protected areas. They also excluded woodlands, marshlands and seed crop agricultural areas without any buffer distance. Significant deviations from the Prior datasets utilized for the LANUV recreation involved LANUV's exclusion of flood plains, for which no Prior was developed, their exclusion of 450 m from exterior areas with residential use, the exclusion of windbreak areas and the specific exclusion of lignite mining sites, as opposed to general mining sites. Additionally, the LANUV study used a number of proprietary or otherwise unavailable datasets that could not be compared to those used to construct the Priors, which included the dataset on settlement- and residential-use areas.

The UBA [59] study investigated onshore wind LE throughout Germany. Once again, typical wind-relevant exclusion constraints were considered, including distance from urban areas, individual dwellings, state and federal motorways, power lines, railways, industrial and commercial areas, camp sites, lakes and rivers and protected areas. Marshlands and forests were also excluded without a buffer distance. Additionally, all slopes above 30 degrees were excluded. Deviation from the utilized Priors was mostly limited to the way criteria were defined; for example, UBA's criterion defined as distances from *individual dwellings* versus the settlement proximity Prior's definition of distances from *urban land coverage*. Most importantly, the UBA study excluded distances from settlement areas up to 600 meters, distances from industrial areas up to 250 meters and distances from campsites up until 900 meters. However, they go on to treat 600-1200 meters from settlement areas, 250-500 meters from industrial areas and 900-2000 meters from campsites as reduced turbine operation zones to prevent exceeding noise limitations. This has been accounted for by increasing the exclusion threshold of each of these priors to a median value. Aside from definition misalignments, the UBA study used criteria definitions very close to those employed in the Priors and, in some cases, the study actually shared the same dataset as well; for example, the CLC [25], as well as the EEA's Nationally Designated Areas (CDDA [87]) and the NATURA2000 [88][4] were all used. Although the UBA study investigated the entirety of Germany, only the results for central Germany (including North Rhine Westphalia, Rhineland Palatine, Hesse, Thuringia and Saxony), as well as those for south Germany (including Bavaria, Baden Wurttemberg and the Saarland) were evaluated here[5].

Sliz [21] investigated onshore wind, open field PV and biomass in central Poland. Exclusion constraints were highly detailed and included distances from settlement area, industrial zones, leisure areas, existing and planned roads, railways, airports, power lines, gas grid, mines, castles, flood areas, lake, rivers, protected areas and forests. Slope, elevation, and aspect were also considered. Similar to the LANUV study, Sliz's exclusion of proximity to flood areas could not be represented using the Prior datasets. Additionally, some of the criteria definitions differed from our own, including distance from *single dwellings* as opposed to settlement areas and *planned* motorways as opposed to existing roadways. Nevertheless, the majority of Sliz's criteria definitions employed were close to our own and also employed the CLC[6], CDDA and NATURA2000 datasets. Only the LE result for onshore wind in Pomorskie is used for the validation.

---

[4] Both CDDA and NATURA200 are incorporated into the WDPA dataset.
[5] The same exclusion constraints and datasets were applied to each region.

Table 4: **Validation results** and utilized thresholds against Prior datasets

| | Constraint | Unit | LANUV [60] | UBA "South" [59] | UBA "Central" [59] | Sliz [21] | McKenna [18] | Latinopoulos [29] | Höltinger "Med" [1] | Höltinger "Min" [1] | Samsatli [20] | Robinius [19] |
|---|---|---|---|---|---|---|---|---|---|---|---|---|
| **Sociopolitical** | Settlement proximity | < m | 600 | 1000 | 1000 | 500 | 800 | 500 | 1200 | 1000 | 500 | |
| | Settlement urban proximity | < m | | | | | | 1000 | | | | 800 |
| | Airfield proximity | < m | 1500 | 1756 | 1756 | | | | | | | 1000 |
| | Airport proximity | < m | 4000 | 5000 | 5000 | 3000 | 1000 | 3000 | 5100 | 5100 | 5000 | 1000 |
| | Roadway main proximity | < m | 40 | 100 | 100 | 150 | 200 | 150 | 300 | 300 | 200 | 200 |
| | Roadway secondary proximity | < m | 40 | 80 | 80 | 100 | 200 | 150 | 300 | 300 | | 200 |
| | Ag. permanent crop proximity | < m | 0 | | | | | 0 | | | | |
| | Railway proximity | < m | 100 | 250 | 250 | 100 | 200 | | 300 | 300 | | 200 |
| | Power line proximity | < m | 100 | 120 | 120 | 200 | | | 250 | 250 | | |
| | Industrial proximity | < m | 0 | 400 | 400 | 250 | 0 | 0 | 300 | 300 | | 300 |
| | Mining proximity | < m | 0 | | | 100 | | 0 | 300 | 300 | | |
| | Camping proximity | < m | | 1500 | 1500 | 450 | | | | | | |
| | Leisure proximity | < m | | | | 450 | | | | | | |
| | Touristic proximity | < m | | | | 1000 | | | | | | |
| **Physical** | Slope threshold | > deg | | 30 | 30 | | 20 | 14 | 8.5 | 11.3 | 8.5 | |
| | Waterbody proximity | < m | 50 | 0 | 0 | 200 | | | | | | |
| | Lake proximity | < m | | | | | | | 1750 | 1000 | | |
| | River proximity | < m | 50 | | | 250 | | 0 | 0 | 0 | 200 | |
| | Woodland proximity | < m | 0 | 0 | 0 | | | | | | 250 | |
| | Wetland proximity | < m | 0 | 0 | 0 | | | 0 | | | | |
| | Elevation threshold | < m | | | | | | | 1750 | 1750 | | |
| **Conservation** | Protected habitat proximity | < m | 300 | | | | 0 | | 0 | 0 | 0 | 0 |
| | Protected wilderness proximity | < m | 0 | | | | | | | | | |
| | Protected biosphere proximity | < m | 0 | 0 | 0 | | 0 | | 0 | 0 | | 0 |
| | Protected bird proximity | < m | 300 | 200 | 200 | | 0 | | 0 | 0 | | 0 |
| | Protected park proximity | < m | 0 | 200 | 200 | 200 | 1000 | 0 | 2000 | 1000 | | 1000 |
| | Protected reserve proximity | < m | 300 | 200 | 200 | 500 | 200 | | | | | 200 |
| | Protected monument proximity | < m | 0 | 500 | 500 | | | | 0 | 0 | | |
| | Protected landscape proximity | < m | 0 | 0 | 0 | 200 | 0 | 1000 | 0 | 0 | | 0 |
| **Econ.** | Windspeed 50m threshold | < m/s | | | | | | 4.5 | | | 5 | |
| | Roadway main proximity | > m | | | | | | | | | 1500 | |
| | Roadway secondary proximity | > m | | | | | | | | | 500 | |
| **Result** | **Reported remaining area** | **%** | **3.3** | **14.1** | **10.2** | **34.8** | **4.9** | **17** | **10.3** | **13.9** | **2.5** | **33.4** |
| | **Computed remaining area** | **%** | **10.9** | **13.9** | **8.8** | **40.4** | **4** | **23.7** | **14.1** | **19.8** | **3.6** | **35.0** |

The McKenna [18] study also investigated onshore wind LE throughout Germany. Distances from settlements, commercial, mixed-building, federal roadways, railways, airports, protected reserves and protected parks were all excluded. Additionally, industrial areas, habitats, landscapes and biospheres were excluded without any buffer zone. On top of this, locations with a terrain slope greater than 20 degrees were excluded. This study also included the use of CLC and OSM [80]in a similar fashion

---

used here[7]. Besides the criterion defined by distances from *mixed building areas*, McKenna's criteria definitions matched very well with those employed for the Prior datasets. Unfortunately, McKenna did not report the raw LE result, and instead the available lands are assigned a suitability factor of between 0 and 1 according to their CLC land cover class. The suitability factor-weighted available lands are then summed together to calculate the total available land. This procedure was recreated using the reported weighting structure; however, as McKenna used a previous version of the CLC compared to that used here, small differences are to be expected. Although McKenna investigated the whole of Germany, only North Rhine-Westphalia was evaluated for validation.

The Latinopoulos [29] study investigated onshore wind in the Kozani region of northern Greece. Using officially mandated exclusion definitions, Latinopoulos excluded distances from protected landscapes, large, small and traditional settlements, roadways, tourism facilities, industrial areas, airports, and archaeological and historical sites. Mining sites, wetlands and irrigated agricultural lands are also excluded without a buffer distance. Additionally, locations characterized by terrain slopes beyond 25% and an average annual wind speed below 4.5 m/s are also excluded. Many of Latinopoulos' criteria definitions differed from the Prior datasets. For example, the exclusion of distances from large settlements (with populations above 2000 individuals), small settlements (populations below 2000 individuals) and settlements otherwise designated as "traditional" differed significantly from the settlement proximity and settlement urban proximity Priors. In addition to this, Latinopoulos excluded distances from archaeological and historical sites for which there is no representation in the Prior datasets. On the other hand, Latinopoulos also used the CLC and NATURA2000 datasets in the same way that other LE researchers have. Latinopoulos' exclusion based on wind speeds inherently also excluded large lakes as well, as these were not defined in the wind speeds dataset used. Therefore, the lake proximity Prior was utilized in the recreation of this study as well, despite not being explicitly indicated.

Höltinger [1] investigated onshore wind LE in Austria for a trio of exclusion constraint sets representing minimal, median and maximal exclusion scenarios. Two validation efforts are derived from this study, corresponding to the minimal and median sets. As exclusions, Höltinger considered the distances from settlement areas, buildings outside settlement areas, roadways, power lines exceeding 110 kV, railways, built-up areas, forests, airports, lakes and rivers, protected areas and major migration routes. Additionally, a maximal slope (8.5° for the median scenario and 11.3° for the minimal) and a maximal elevation defined at the alpine forest line are enforced. Höltinger used many of the same datasets used here, including CLC, OSM, CDDA and NATURA2000, although many of the criteria definitions used here differ significantly from the Priors, alongside the use of several Austria-specific datasets. Most notably, Höltinger's exclusion of animal migration routes and buildings outside settlement areas are completely unrepresented in the Prior datasets. Additionally, Höltinger employed many of the conservation-relevant exclusions on a case-by-case basis, while the Prior datasets cannot distinguish between individual designations. For this validation, only the eastern region of Burgenland is recreated.

Samsatli [20] investigated onshore wind LE in the United Kingdom. Ten exclusion constraints were considered in total, including distances from developed land, roadways, airports, rivers, woodlands, protected areas and power lines. Sites of special scientific interest were also excluded, along with slopes exceeding 15% and average annual wind speeds below 5 m/s at 45m. Samsatli included constraints based off grid connection and accessibility distance, which were expressed as a minimal distance from major roadways (for grid connection) and major and secondary roadways (for accessibility), which explains the use of the roadway Priors in the economic group of Table 4. Along with these, Samsatli also excluded all locations within five times the rotor diameter of existing turbines. The definition of distances from *developed land* differed from the settlement proximity Prior's definition, and there was no Prior created to represent distances from preexisting turbines. Furthermore, when considering protected areas, Samsatli only excluded designations of "Sites of Special Scientific Interest" (SSSI), and while these sites are included in the WDPA dataset and therefore also in the various conservation group Priors, they are not specifically selected by any of the conservation group criteria definitions. The protected habitat proximity was found to best match with the SSSIs, however, which is why it was used for validation over the other options. Instead of evaluating the entirety of the UK, only the southwestern portion of England[8] was used for validation.

As the final validation study, Robinius [19] also investigate land eligibility for wind turbines in Germany. As the investigator of this study is also an author of this work, this study offered a unique opportunity to compare the operation of the described framework to a conventionally-evaluated LE study where the exact datasets and procedures are known. In this study buffered regions around settlement areas, all airports, roads, railways, industrial areas, and protected parks are all excluded. Furthermore, protected bird zones, landscapes, biospheres, and habitats are excluded without any buffering. Robinius' choice of protected areas differs from the way they are used in the Prior datasets, nevertheless the CDDA and NATURA2000 datasets are used so the differences are expected to be minimal. Robinius also made use of the CLC and OSM datasets, however in both cases a previous version is used compared to those used to produce the Priors. Once again, instead of recreating the entire study

---

[7] A previous OSM dataset extract and previous CLC version was used.

[8] Composed of the Cornwall, Devon, Somerset and Dorset regions

## Höltinger Replication

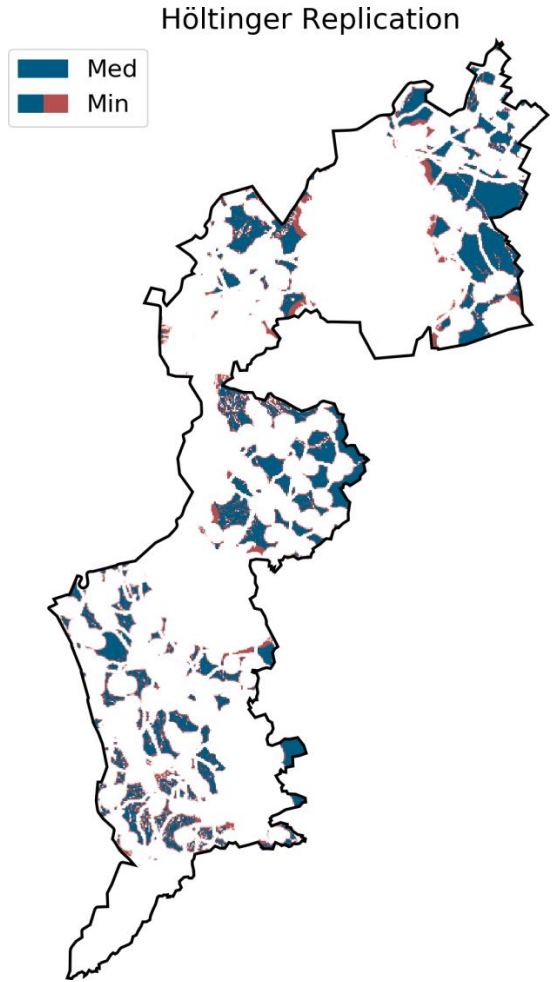

*Figure 4: **Validation recreation** results of Höltinger's [1] exclusions for the region of Burgenland, Austria. Colored regions imply eligibility*

area only the south-west portion of Germany[9] is chosen for validation.

### 4.2. Validation Results

Replication results were the least successful with the LANUV report, where a 7.6% difference was observed from the reported left over area. As described above, there were several discrepancies between the datasets and criteria used for the Prior datasets versus those that were employed by LANUV. Most notable, however, was the outcome after applying the constraints related to settlements. After subtracting areas within 600 m of general settlements and within 450 m of rural-use buildings, LANUV reports that 22% of the state's area remains. Comparing this to a value computed by the described framework, after applying a 600 m buffer to the settlement proximity Prior (which also excludes rural settlements), 55% of the state's areas remains. This indicates that the dataset used for the LANUV study differs extensively from the urban area designations in the CLC dataset. Despite the claim that their sources are openly available, they were only found to be obtainable in a form of a web-gui, which could not be extracted and used within our framework. This outcome does not serve to validate or invalidate either implementation, but rather serves as a reinforcement of the point that the use of open and consistent data sources is necessary for broad context LE analyses, as well as for those analyses that depend on them.

Following the LANUV study, the Latinopoulos replication showed the second to worst result, with a 6.7% deviation in the total remaining land. Just as with the LANUV report, several differences from the Prior datasets existed. In this case, key considerations (archaeological and historical sites) could not be included in the replication, and as such it follows that less land would be excluded. In spite of this, it seems that the most likely reason for the discrepancy observed is Latinopoulos' wind speed constraint. Using the same wind speed dataset [89] used by Latinopoulos, it was found that excluding all locations with an average wind speed below 4.5 m/s resulted in only 9% remaining area before considering any other constraints as well. Clearly, this result does not concur with Latinopoulos' reported result and therefore another operation that was not detailed in the publication must have been performed on the wind speeds prior to exclusion, otherwise Latinopoulos must have used a previous version of this dataset[10]. In any case, without further information regarding Latinopoulos' procedure, this discrepancy could not be investigated further.

After these, the Höltinger replications resulted in a 3.8% deviation for the median scenario and a 5.9% deviation for the minimal scenario. Once again, some criteria are missing (animal migration routes) and differing criterion definitions (single dwelling) are present in these recreations. In this case, it was not possible to identify a single dataset as the main cause of the observed discrepancies. Despite these differences, however, consistency between Höltinger's study and the recreation can be confirmed via visual comparison of the two results. For this purpose, the results of the Höltinger replication are shown in Figure 4. A comparison between the two replicated scenarios illuminates their difference, where a significantly smaller deviation from the reported value is found for the median scenario compared to the minimal. One way to interpret this result would be that as other exclusion constraints become more confining (for example, the increased exclusion range around settlements), the areas that would have been excluded by the missing constraints are excluded anyway, because they have a higher likelihood of overlapping with the constraints that are included. This dynamic of overlapping constraints presents an interesting investigation point that will be explored in the following section.

The Silz replication showed a 5.6% deviation for reasons similar to those discussed above. The replication did not include the original exclusion of flood plains, and two criteria definitions differed significantly. However, the resulting discrepancy was less than that of the previous three studies, and so is not discussed in further detail.

---

[9] Composed of Saarland and Baden Württemberg

[10] Although no indication of another version of the dataset could be found

The four remaining studies, UBA, Samsatli, McKenna and Robinius were found to match very well with -0.2% and -1.4%, 1.1%, 0.9%, and 1.6% deviations respectively. These studies also showed interesting discrepancies, although these were clearly found to have insignificant impacts on the final results. The Samsatli study, for instance, did not include the original exclusion of distances from preexisting wind turbines. Similarly, the UBA study proximity exclusion from single dwellings did not precisely match the settlement proximity Prior used in the replication. Nevertheless, these studies showed general agreement in the datasets used compared to the Prior datasets and their criteria definitions were more often than not a match to our own. Therefore, the significant agreement of these studies with the reported results suggests that the described framework is indeed operating as expected and, furthermore, that the Prior datasets are sufficient for conducting generic LE analyses. Furthermore, these results suggest that if complete knowledge was held concerning the datasets and practices used by investigators of the previously discussed studies, these results could also be replicated with improved accuracy. Taking into consideration that the replication studies with the largest deviations (LANUV and Latinopoulos) were associated with irreconcilable issues with a particular dataset, that the other replications with relatively large deviations seemed to be missing significant criteria and because the studies that were not subject to these issues (at least not where key criteria were concerned) showed very strong agreement, it can be finally concluded that the described framework is validated and reliable[11].

## 5. Evaluation

When evaluating LE using multiple exclusion constraints, it logically follows that some locations are redundantly excluded by more than one constraint. From this, the question arises as to how much of a factor this dynamic is. Following this line of thought, it can be concluded that certain constraints should be relatively more valuable than others for a variety of reasons. Of course, a constraint that excludes a large portion of the land is a valuable consideration; however, so too is a constraint that excludes a smaller proportion of locations where those excluded are unique to that constraint. Similarly, a constraint that has a high tendency to overlap the excluded areas of another constraint could also be considered important because it reduces the need for possessing detailed data on the overlapped constraint (as having such data would not significantly change the LE result). For the purposes of this discussion, these measures of constraint value will be called the *independent impact*, the *exclusivity* and the *overlap*.

The Höltinger replication discussed in the previous section provides excellent insight into this dynamic of constraint value. In this replication, a general agreement was apparent between our modeled result and Höltinger's reported result, both visually and with regard to the total percentage of remaining land. Our results were found to exclude 5.9% less land than Höltinger in the minimal exclusion scenario, however. This makes sense though, as two key constraints used by Höltinger, namely proximity to animal migration routes and proximity to single dwellings, were not included in our validation. Although the influence of these two constraints could not be investigated in detail, it is expected that these two constraints possess a relatively small impact compared to Höltinger's other constraints, yet nevertheless have high incidence in remote areas and are therefore valuable in regard to the exclusivity of their exclusions. However, it is also expected that animal migration routes are likely to link designated habitats, and furthermore that the routes themselves are likely to follow close to lakes and rivers; all of which are also considered as exclusion constraints in Höltinger's study as well as in our validation. With regard to single dwellings, although there are sure to be such dwellings in remote areas, it is also expected that a higher proportion of these are found closer to settlements which, once again, are also excluded in both Höltinger's study as well as in our validation. Considering that it is not only these features themselves that are excluded in the LE analysis, but also all areas within a given buffer distance, it is therefore plausible that much of the areas that would have been excluded by the two constraints which were missing in our validation were already excluded by the other constraints which were included. It also stands to reason that in a more restrictive exclusion scenario utilizing the same criteria definitions, a greater portion of the missing-exclusions are overlapped by the same mechanism, and therefore the difference between a value given by Höltinger and a recreated result would be closer. This is exactly what is seen in Höltinger's median exclusion scenario where, amongst other changes, the excluded distance from settlements is increased from 1 to 1.2 km, the excluded distance from lakes is increased from 1 to 1.7 km and the excluded distance from protected parks is increased from 1 to 2 km. Between these two scenarios, the difference between the reported and replicated result decreases from 5.9% to 3.8%.

---

[11]This is not to say that all of the Prior datasets are valid for all areas. Complete validation of the underlying datasets (such as the OSM and WDPA) is another matter that extends far beyond the scope of the work discussed here

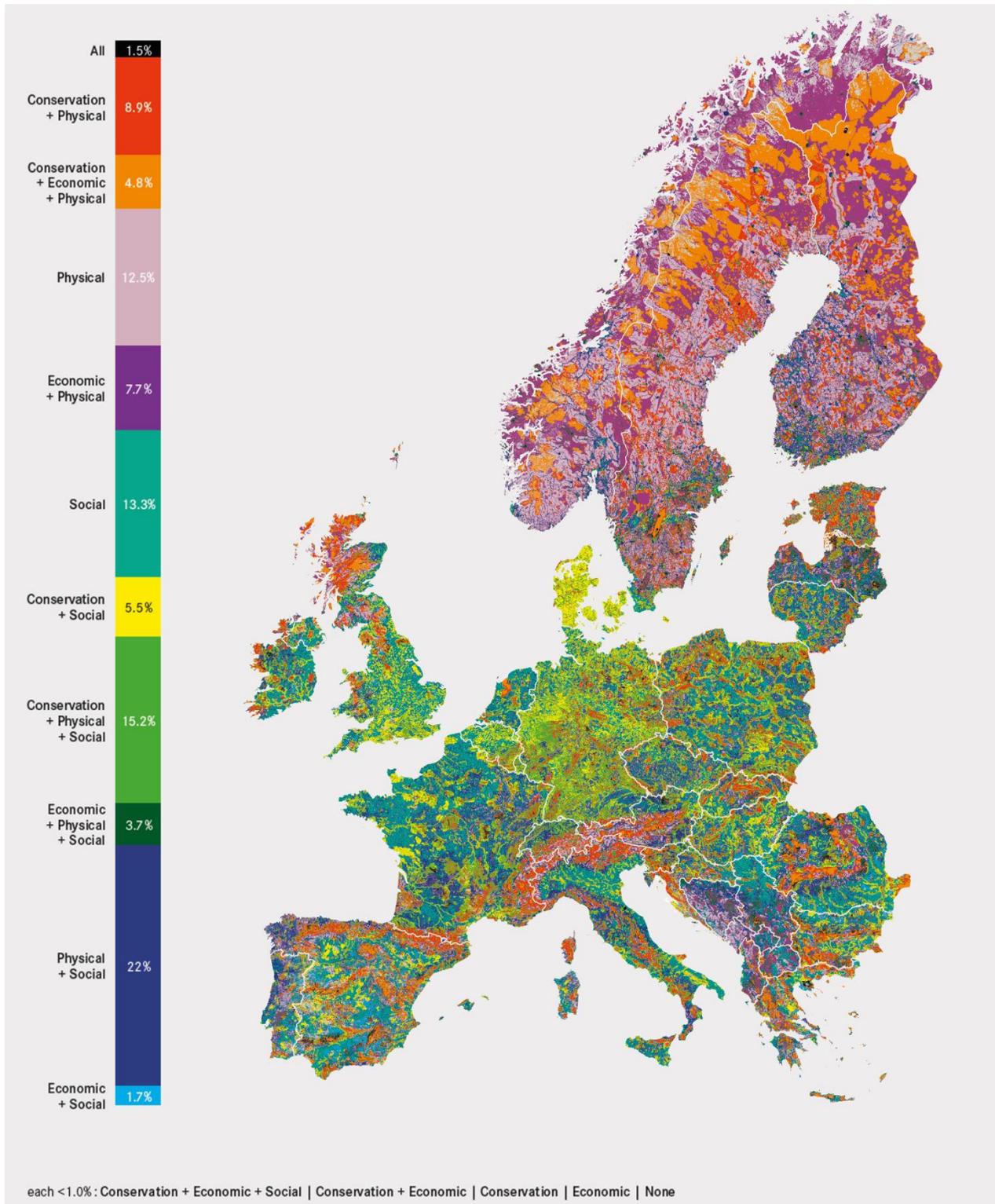

**Figure 5: Motivational Contributions** *of the four motivation groups to the typical exclusions across Europe.*

Thus, how valuable are the missing constraints in the Höltinger replication in comparison to the constraints that are included? When is it important to consider a particular constraint with high detail in an LE analysis and, on the other hand, when is it expected that including a particular constraint will result in little to no change in the final result? The following sections detail the effort to illuminate this tendency by using the Prior datasets and, in turn, to determine the relative importance of the various constraints according to the previously mentioned measures. This is accomplished in several stages. First, a spatial intuition for where constraints enact their exclusions is developed by directly plotting the aggregated motivation groups and observing where they tend to overlap. Following this, the independent impact of each constraint is measured at the European level, as well as for each country included in the Prior datasets. Finally, by comparing the constraints to each other one to one, their exclusivity and overlap is measured again for Europe, as well as at the country level. The section then concludes by discussing the outcomes of these investigations.

*5.1. Constraint Mapping*

For each of the Prior datasets shown in Table 3,[12] the associated typical exclusion thresholds from Table 2 were independently evaluated across the entire European continent. Instead of plotting each of these results, however, these independent exclusion results were aggregated according to their motivation groups whereby, if any one pixel were excluded by at least one constraint, it would remain excluded in the final, aggregated result. Finally, these plots were overlaid with one another, with a value assigned to each pixel according to the combination of the four motivation groups contributing to the exclusion of that pixel. The resulting map expresses 16 possible combinations, ranging from indicating that no motivation groups excluded the considered pixel to indicating locations where all motivation groups excluded the considered pixel. **Figure 5** displays the result of this effort. This figure is not intended to reflect the exclusions for any particular technology, or in any particular area, but nevertheless offers an understanding of where motivation groups tend to play a dominant role in LE analyses[13].

By themselves, it is seen in **Figure 5** that physical exclusions collectively impact 77% of all of the area in Europe, sociopolitical exclusions impact 63%, conservation exclusions impact 36%, and economic exclusions impact 19%. Regarding overlap between the motivation groups, the first observation to note is that all 16 combinations are observed somewhere in Europe, although some combinations are expressed far more frequently than others. The largest shares are found in the combination of physical plus sociopolitical exclusions (22% of pixels), conservation plus physical plus sociopolitical exclusions (15.2%), sociopolitical exclusions (13.3%) and physical exclusions (12.5%). Interestingly, about 1.5% of land is impacted by all four motivation groups, while slightly less than 1% is not impacted by any group.

Beyond summary quantities, it can be seen that, despite the broad inclusion of many criteria in each motivation group, a very strong spatial dependence remains. Denmark and Germany, for example, are almost entirely affected by conservation, as well as sociopolitical exclusions, while the same considerations play a minor role in the Nordic countries, where economic and physical concerns are prominent. Spain appears to have a similar distribution to

Romania and Bulgaria, where mountainous ranges lead to conservation, economic and physically-motivated exclusions surrounded by large areas where sociopolitical exclusion areas dominate. More precisely, there is extensive variation in motivational contribution within countries. Although not dominant, Germany also exhibits a dependence on physical constraints, although these appear to have a non-uniform distribution across the countryside and are particularly present in the south. Furthermore, Switzerland is almost entirely covered by the physical exclusion group, but has pockets of conservation-related exclusions and is neatly bisected into a northern and southern region where sociopolitical considerations are also impactful. France and the UK can both be seen to transition between areas where sociopolitical based exclusions play the major role to areas where physical and conservation considerations also become important.

Although there is quite a bit of structure shown in **Figure 5**, it is also limited in the amount of meaningful information it can offer. The most important conclusion that can be drawn from this figure, however, is that even at a very high level of aggregation, the spatial dependence of LE constraints remain more or less chaotic. From this, the conclusion can be drawn that general simplifications and tenets regarding LE behavior, such as "including $X$ constraint always results in a $Y$% reduction in the available land," are not substantiated across large scales or even, in many cases, within countries.

*5.2. Independent Impact*

To determine independent impacts, the Prior datasets were again evaluated with the typical thresholds and the percentage of land each constraint excluded was recorded for all countries in the study area, as well as for the entire European study area. Figure 6 shows the result of this in the form of a heat map wherein the constraints have been ordered from left to right in the order of their average exclusion percent across all nations[14]. As per the previous discussions, this figure is not intended to suggest that a given constraint will always exclude the reported percentage of area within these countries, as this percentage depends heavily on the threshold used in each case and, especially when a sub-region is investigated, on the spatial variability within these countries. Instead, the figure is merely meant to rank the relative independent contributions of each constraint as shown by the left to right order. Therefore, the resulting order comments on the relative *impact* value of each constraint, meaning that constraints further towards the left are more important in the sense that, when their consideration is warranted for the technology in mind, they tend to exclude the most land.

It can be seen from the figure that woodland and agriculture proximity, both in the general sense, typically have the greatest impact across Europe (excluding 51% and 50%, respectively). Following these are, for example,

---

## Exclusion

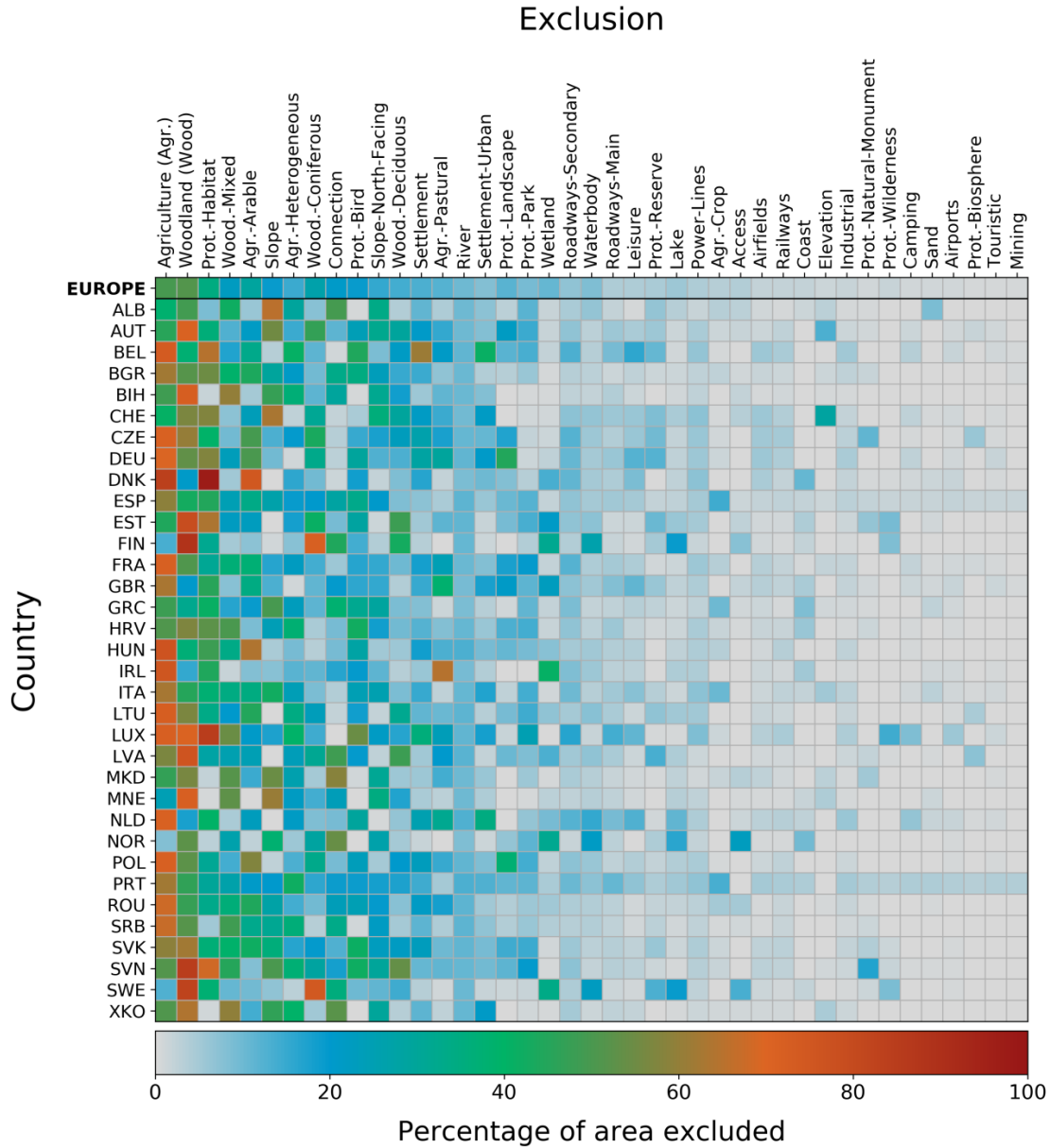

***Figure 6: Independent contributions*** *of constraints as determined by each of the Prior datasets.*

protected habitat proximity (35%), connection distance (21%) and slope threshold (19%). Here, it is also seen that certain subgroups of agricultural and woodland proximity have a larger impact than others; for example, agriculturally arable proximity is first in the agriculture group with 29% exclusion across Europe and coniferous woodland leads the woodland group with 27%. At the national level, : again emphasizes the point that the impacts of various constraints depend heavily on the region in question, thereby discouraging generalizations at the international scale. River-proximity represents the only legitimate exception to this conclusion, where the percentage exclusions are consistently observed between 9% and 13% for all countries. Access-distance appears to have a small impact for all countries, except in the Nordic regions, suggesting that Europe's road network is approximately comprehensive. There are, of course, constraints with a consistently low impact, such as mining-proximity, touristic-proximity and protected biosphere proximity, although these results should not be interpreted as these criteria leading to uniformly low exclusions everywhere. Despite the large exclusion buffer of 5 km, airport-proximity results in very little total land exclusion (averaging only 0.92%), with the exception of Luxembourg, where a figure of 5.7% is seen. Slope-threshold is an extreme example of regional variability, ranging from 0.04% in the Netherlands to 65% in Albania. It can also be seen from : that the high conservation exclusions seen in Denmark results almost entirely from protected habitats (97%). By comparison, protected habitats are slightly less impactful in Germany; however, Germany is also heavily affected by protected landscapes that together cover the majority of the country, as observed in Figure 5.

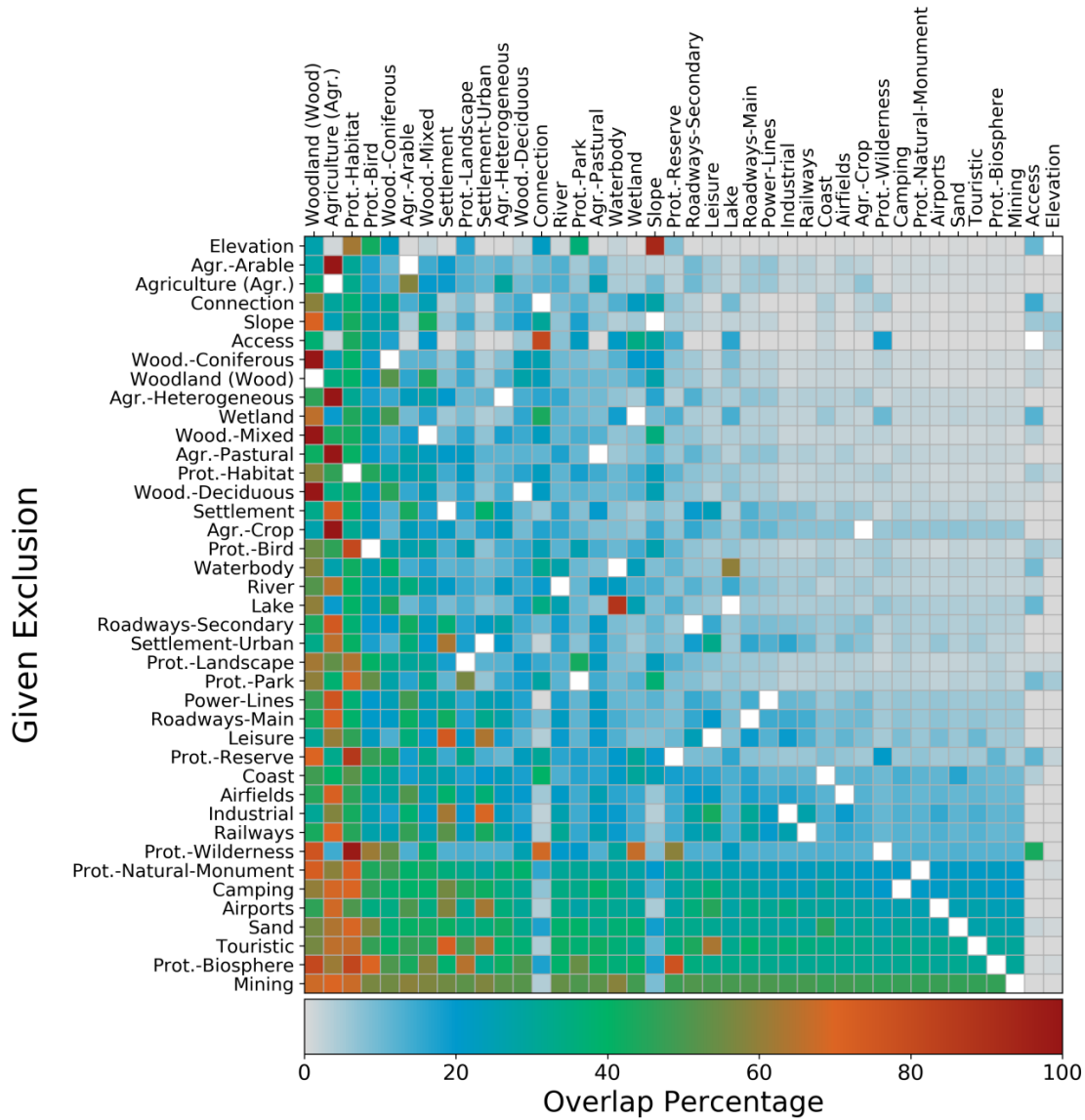

***Figure 7: Overlapping and exclusivity contributions*** *by measuring the percent overlap between constraints at the European level.*

### 5.3. Overlapping and Exclusivity Impact

Figure 6 displays the independent impact of each constraint in its typical expression, but does not communicate any information regarding how these constraints overlap one another. This dynamic is investigated by considering each constraint computed, using the same procedure as in the previous steps, and determining the extent to which the given constraint overlaps with that of another, overlapping, constraint. This is performed for every pairwise combination of constraints, with the results shown in Figure 7. Similar to Figure 6, the order of constrains in both the *given* and *overlapping* dimensions address their relative importance. The *given* dimension's order is found by determining the average of all overlapping percentages for the given constraint and is shown top-down, in increasing order. Constraints near the top of this figure tend to have little overlap with the other constraints and therefore have a higher measure of *exclusivity*, meaning that they are important to properly exclude since the areas excluded by these constraints are less likely to also be excluded by other constraints. On the other hand, the *overlapping* dimension's order is found by determining the average percentage overlap of a constraint with respect to all given constraints, and is shown as decreasing from left to right. In this way, constraints towards the left tend to have a higher measure of *overlap* since these have a tendency to exclude areas that a researcher may want to exclude for other reasons (and potentially may not have accurate data for).

Some relationships expressed in Figure 7 are expected and serve to validate the process from a logical consistency perspective. For instance, when given an arable agricultural constraint, the resulting exclusions are completely overlapped by a general agricultural constraint. Similarly, even though the lake proximity and waterbody proximity are derived from different data sources, lake proximity is completely overlapped by the waterbody proximity, yet the inverse is not valid as the waterbody proximity also includes large rivers. Other relationships are not as obvious, yet are nevertheless plausible, such as after having excluded general settlement-buffered areas, large portions of areas that would have been excluded by proximities to leisure, industrial, airports, camping, touristic and mining areas are already excluded. Yet, once again, the opposite is not valid, as these constraints have relatively little overlap with the whole area excluded by the settlement proximity constraint. Furthermore, the overlapping nature between the protected area definitions, which was mentioned in Section 3.3, is shown in detail. Naturally, constraints that have large independent impacts also tend to rank high in both the exclusive and overlapping measures, so it is clear that these constraint measures are not fully independent of each other. For example, the general woodland and agriculture proximities have an unfair advantage compared to the other constraints in the manner that the overlapping ranking (left to right) was computed, as these constraints have multiple sub-constraints with which they overlap completely. Connection distance and slope threshold are also notable in this way, considering that they maintain a high or at least median-to-high score in all three measures. Not all constraints maintain such a semi-constant position, however. Both the elevation threshold and access distance constraints, for instance, have a mid-to-low impact, are last in terms of overlap, yet are seen to rank first and sixth in terms of exclusivity. These two exclusions only overlap each other around 15-20% as well, so the areas they exclude are generally distinct.

*5.4. Discussion*

Regarding the individual constraints, it is apparent that several constraints score low in all three measures; including touristic, camping, mining, biospheres and airport proximity. Airport proximity is a particularly interesting member of this group, considering that it is an exclusion considered in more than half of all the reviewed LE analyses. Although also scoring mid-to-low in all measures, airfield proximity is also consistently higher ranked than airport proximity. Likewise, there are several constraints that are routinely highly valuable in all three measures, including those related to distances from agricultural, woodland, settlement and habitat areas, as well as connection distance. Some very commonly included constraints, for example roadway, railway, power line, lake and river proximity, consistently rank somewhere in the middle, suggesting that these considerations are valuable considerations but are nevertheless unlikely to play a dominant role in an LE result. Lastly, a few constraints, such as access distance and elevation, span a range of measures, as previously discussed. In particular due to their exclusivity score, these considerations are valuable, as their exclusions will tend to be unique.

The evaluations performed in this section are presented to give LE researchers a general clue regarding where and when to consider certain criteria. Furthermore, although the constraints are evaluated specifically at their typical threshold value, these analyses also provide insight into how the underlying criteria interact with one another in any situation (such as when expressed as decision factors in MCDM analyses). In any case, the point is reiterated at nearly every stage that generalizations across large spatial scales are not substantiated, as the impact of any one constraint depends heavily on where the evaluation is taking place and on which other constraints are considered as well. More specifically, Figure 7 is particularly useful, in that it shows how consideration of a particular constraint can diminish the importance of another. For example, if one does not have detailed information regarding leisure, touristic and camping areas, they can generally assume that a generous exclusion of areas surrounding settlement areas will cover the missing constraints anyway. Although not presented at the regional level, this overlapping dynamic is certain to be sensitive to geographic areas as well, so such an argument should only be used as a last resort when detailed datasets are unavailable.

As a final note, it is easy to see how an inconsistent treatment of exclusion constraints in LE analyses as discussed in Section 2, on top of the already highly complex nature of their interactions, very quickly leads to the incompatibility and disparate nature of LE analyses currently observed in the literature. Above all, this supports the notion that a standardized LE approach such as that presented in this work is a necessity for the LE community, as well as for those who depend on the results of LE analyses.

6. CONCLUSION

The discussion herein has explored the general concept of land eligibility (LE). LE is described and it is pointed out how it is a common analysis applied to a wide variety of RES technologies in any geographic area, and typically precedes more complex analyses, such as energy system models. But as LE and LE-subsequent analyses grow in spatial scope to national and even international scales, the need for consistency in the LE approach becomes increasingly important. However, when trying to make determinations in these spatial contexts, the current state of LE in the literature is insufficient due to a multitude of inconsistencies between studies: inconsistent criteria definitions, inconsistent or otherwise opaque methodologies and inconsistent dataset usage. The novel work described here discusses several efforts to alleviate these issues in the form of presenting a general framework by which LE investigations can be performed.

Regarding criteria definitions, 55 LE analyses covering wind, PV and other RES technologies were reviewed in order to formulate an understanding of the criteria that routinely arise in LE analyses. Although each study considered its own set of criteria for its relevant technology, defined in each case more-or-less at the researcher's discretion, the myriad of definitions were generalized into a set of 28 independent criteria (and additional sub-criteria) and organized into one of four groups according to their underlying motivation. Additionally, a range of threshold values observed when these criteria are expressed as exclusion constraints within the literature is provided for each criterion. Following this, a fully transparent methodology for performing LE analyses was proposed, which is capable of operating in any geographic area and is not limited in spatial scope. Moreover, the described model is capable of manipulating an extensive variety of vector and raster source formats, meaning any collection of underlying data sources can be incorporated in a single analysis. The model is implemented in the Python programming language, has only minimal dependencies (all of which are open-source) and is made freely available under the GitHub project GLAES (Geo-spatial Land Analyses for Energy Systems) [47]. Finally, using various publically-available data sources with broad spatial contexts, a collection of standardized datasets have been produced for the European continent that realize the majority of the generalized criteria identified previously. In all 45 of these standardized datasets, referred to as *Priors*, were produced and can be used for rapid LE analyses in Europe, either by themselves or in conjunction with other geospatial datasets.

The framework detailed in the methodology section makes consistent evaluations of LE within the European context readily available, although the question remained as to whether these evaluations show agreement with other LE evaluations found in the literature. Therefore, eight studies were identified for recreation, of which four showed very strong agreement (within a few percentage points of the reported resulting available land) and the remaining four showed agreement but with significant differences. Interestingly, this agreement was reached despite incomplete information and differing criteria definitions. For the studies that did not match well, the discrepancies are understood to originate from one or two specific considerations made by the original authors. In all cases, attempts at amending these replications using the authors' original methods and datasets were limited by the use of a proprietary datasets or unclear methodologies. From these results, it can be concluded that the described framework is validated. Moreover, the divergence from some of the replication studies does not invalidate either approach but rather reinforces the importance of consistent methodologies and open data practices.

Following this, general LE dynamics were investigated to build intuition on how the various criteria behave in relation to one another in the European context. First, the four motivation groups were visualized across Europe and it was observed that, across the continent, a strong spatial dependence is present, despite a high degree of aggregation. From this, it was concluded that it is not advantageous to generalize LE in large spatial contexts; once again reinforcing the use of a standardized LE approach. Following this, the various constraints were relationally ranked according to their independent impact, the exclusiveness of their excluded areas and their tendency to overlap the other constraints. It was found that some constraints rank highly in all three measures, such as woodland and agriculture proximity. Meanwhile, others tend to rank low in all three measures: for example, proximity to biospheres, touristic, leisure, camping areas and, most notably, airports. Still other constraints possessed a mixture of rankings, such as elevation thresholds and maximal access distances.

This work sets the stage for a variety of future efforts. For one, the creation of the Prior datasets makes LE analyses in Europe easy and quick to evaluate. Therefore, an effort could be made by researchers and policymakers to define region-specific (preferably subnational) exclusion sets for various technologies in their area. Such an effort could benefit the entire RES community on several levels – it could serve to identify new criteria that are not already incorporated as Prior datasets, it could help validate the underlying datasets as locals in these regions would be more able to identify missing or misrepresented features and it could serve to make energy system design and other such analyses more sensitive to realistic local preferences for RES installation. Usage of the Priors also theoretically makes MCDM possible anywhere on the European continent, for which regional characterizations can also be made. For areas that possess equivalent data to that used for the Prior datasets, such as potentially in North America or China, it would be interesting to investigate the interdependence of these criteria in these contexts and compare the findings against those that were discussed here. Progress in these areas can also serve to hone data collection efforts in developing countries and other regions for which data is currently sparse. In any case, beyond the future use of the described framework, it is the authors' hope that the work presented here further invigorates individuals, research groups and governmental organizations across the world to participate in open data practices and continually strive for consistency wherever possible.

# Appendix A: Criteria Consideration References

| Group / Criteria / *Sub-Criteria* | EEA [41] | Clifton [50] | LANUV [60] | UBA [59] | Vandenbergh [73] | DLR [71] | Babani[45] | Krewitt [58] | Hansen [55] | Ma [62] | Rodman [37] | Bennui [49] | Tegou [22] | Ummel [72] | Lejeuna [61] | Ramirez [68] | Gastli [53] | Janke (wind) [57] |
|---|---|---|---|---|---|---|---|---|---|---|---|---|---|---|---|---|---|---|
| **Sociopolitical** | | | | | | | | | | | | | | | | | | |
| Settlements | | | X | | | | | X | X | | | | | X | | X | | X |
| *Rural* | | | X | X | | X | | | | X | | X | X | | X | | | |
| *Urban* | | | X | X | X | | | | | X | X | X | X | | X | | | |
| Roads | | | | | | | X | X | X | | | | | | X | | | |
| *Main* | | | X | X | | | | | | | | X | | | | | | |
| *Secondary* | | | X | X | | | | | | | | | | | | | | |
| Airports | | | | | | | | X | X | | | X | | | | X | | |
| *Airports* | | | X | | | | | | | | | | | | X | | | |
| *Airfields* | | | X | | | | | | | | | | | | X | | | |
| Power Lines | | | X | X | | | | | X | X | | | | | X | | | |
| Historical Sites | | X | | | | | X | | X | | | X | X | | | | | |
| Railways | | | X | X | | | X | X | | | | | | | X | | | |
| Agriculture | | | X | X | | X | X | | | X | X | | | | X | | | X |
| Industrial Areas | | | | X | | | | | | | | | | | X | | | |
| Mining sites | | | X | | | | | | | | | | | | | | | |
| Recreational Areas | | | | | | | | | | | X | | | | X | | | |
| Camp sites | | | | X | | | | | | | | | | | | | | |
| Tourist sites | | | | | | | | | | | | X | | | | | | |
| Radio Towers | | | | | | | | | | | | | | | X | X | | |
| Gas Network | | | | | | | | | X | X | | | | | | | | |
| Power Plants | | | | | | | | | | X | | | | | | | | |
| **Physical** | | | | | | | | | | | | | | | | | | |
| Water Bodies | | X | | | | | X | X | | X | | X | X | X | | | X | |
| *Lakes* | | | X | X | | | | | | X | X | X | | | | | | |
| *Rivers* | | | X | X | | | | | | X | X | X | | | | | | |
| Wetlands | | | X | X | | X | | X | X | | | X | | | | | | |
| Slope | | X | X | X | | X | | | | X | X | | | | | | X | |
| Woodlands | | | X | X | | X | X | X | X | X | | | X | | X | | | |
| *Coniferous* | | | X | | | | | | | | | | | | | | | |
| *Deciduous* | | | X | | | | | | | | | | | | | | | |
| Vegetation | | | | | | | | | | | X | | | | | | | X |
| Land Instability | | | | | | | | | | | | | | | | | | |
| *Earth quake* | | | | | | | | | | | | | | | | | | |
| *Flood plains* | | | X | | | | | | | X | | | | | | | | |
| *Land slide* | | | | | | | | | | | | | | | X | | | |
| Elevation | | | | | | | X | | | | X | X | | | | | | |
| Soil Composition | | | | | X | | | | | | | | | | X | | | |
| Aspect | | | | | | | X | | | | | | | | | | | |
| **Conservation** | | | | | | | | | | | | | | | | | | |
| Protected Areas | X | | | | | | X | X | | | | | | X | X | X | | X |
| *Parks* | | | X | X | | | | | | | | | | | X | X | | |
| *Reserves* | | | X | X | | | | | | | | | | | X | | | |
| *Landscapes* | | | X | X | | | | | | | | | | | X | | | |
| *Natural Monuments* | | | X | | | | | | | | | | | | X | | | |
| *Wilderness* | | | X | | | | | | | | | | | | | | | |
| Protected FFH | X | X | | | | | X | X | X | X | | | | X | X | X | | |
| *Habitats* | | | X | X | | | | | X | | X | | | | | | | |
| *Biospheres* | | | X | X | | | | | | | | | | | | | | |
| **Economic** | | | | | | | | | | | | | | | | | | |
| Resource | | | | | | | | | | | | | | | | | | |
| *Wind speed* | X | | | | | | X | X | X | | X | X | X | | | X | | X |
| *Irradiance* | | X | | | X | X | | | | | | | | X | | | | |
| Material transport | | | X | | | X | X | | | | X | | X | X | | X | | X |
| Connection cost | | | X | | X | X | X | | | | X | | X | X | | X | | X |
| Land Value | | | | | | | | | | | | | X | | X | | | |

| | Janke (PV) [57] | Tegou [70] | Funabashi | Lehman[39] | Phuangpornpitaka [67] | Haaren[24] | Zhou[75] | AlYahyai[48] | Grassi[54] | Ouanmni[66] | Sultana[69] | Aydin[26] | Aydin[26] | Gassi[52] | Gorsevski[46] | Sliz (Wind)[21] | Sliz (PV) [21] | Sliz (Biomass) [21] |
|---|---|---|---|---|---|---|---|---|---|---|---|---|---|---|---|---|---|---|
| **Sociopolitical** | | | | | | | | | | | | | | | | | | |
| Settlements | X | X | | | | | X | | X | | | | X | X | X | X | | X |
| *Rural* | | X | | | | X | | | | | | X | X | | | X | | |
| *Urban* | | X | | X | X | X | | X | | | X | X | X | | | | X | |
| Roads | | | | | | X | X | X | | | X | | | X | | X | | X |
| *Main* | | | | | | | | | X | | | | | | | X | X | |
| *Secondary* | | | | | | | | | X | | | | | | | | | |
| Airports | | X | | X | | | X | | X | X | X | X | X | X | X | X | | |
| *Airports* | | | | | | | | | | | | | | | | | | |
| *Airfields* | | | | | | | | | | | | | | | | | | |
| Power Lines | | | | X | | | | | | | X | | | | | X | | |
| Historical Sites | | X | | | | | | | | | | | | | | X | | |
| Railways | | | | | | | X | | X | | | | | | X | X | X | X |
| Agriculture | X | X | | X | | | | | X | | | | X | X | X | X | X | |
| Industrial Areas | | | | | | X | | | | | X | | | | | X | | |
| Mining sites | | | | | | | | | | | X | | | | | X | X | |
| Recreational Areas | | | | | | | | | | | X | | | | | X | | |
| Camp sites | | | | | | | | | | | | | | | | X | | |
| Tourist sites | | | | | | | | | | | | | | | | | | |
| Radio Towers | | | | | | | | | | | | | | | | | | |
| Gas Network | | | | X | | | | | | | X | | | | | | | |
| Power Plants | | | | | | | | | | | X | | | | | | | |
| **Physical** | | | | | | | | | | | | | | | | | | |
| Water Bodies | | | | X | | | | | X | | X | | X | | X | | | |
| *Lakes* | | | | | | X | X | | | | X | X | X | | | | | |
| *Rivers* | | | | | X | | X | | | | X | | X | | | X | | |
| Wetlands | | X | | | X | | | | | | | X | X | | X | | X | |
| Slope | | X | X | | | X | X | X | X | X | X | | | X | | | X | |
| Woodlands | | | | | X | | X | | X | | | | | | | X | X | X |
| *Coniferous* | | | | | | | | | | | | | | | | | | |
| *Deciduous* | | | | | | | | | | | | | | | | | | |
| Vegetation | X | X | | | | X | | | | | X | | | | | | X | |
| Land Instability | | | | | | | | | | | | | | | | | | |
| *Earth quake* | | | | | | | | | | | | | | | | | | |
| *Flood plains* | | | | X | | | | | | | | | | | | X | X | X |
| *Land slide* | | | | | | | | | | | X | | | | | | | |
| Elevation | | | | X | | | X | | | | | | | | X | | | |
| Soil Composition | | | | | | X | | X | | | | | | | X | | X | |
| Aspect | | | | | | | | | | | | | | | | | X | |
| **Conservation** | | | | | | | | | | | | | | | | | | |
| Protected Areas | X | X | | | | | X | | X | X | | | | | | X | X | X |
| *Parks* | | | | X | | | | | | | | | | | | X | X | X |
| *Reserves* | | | | | | | | | | | X | | | | | X | X | X |
| *Landscapes* | | | | | | | | | | | | | | | | X | X | X |
| *Natural Monuments* | | | | | | | | | | | | | | | | | | |
| *Wilderness* | | | | X | | | | | | | | | | | | | | |
| Protected FFH | | X | | X | | | X | | X | X | X | | | X | | X | X | X |
| *Habitats* | | | | X | | X | | X | X | | | X | X | X | | X | X | |
| *Biospheres* | | | | X | | | | | | | | | | | | X | X | |
| **Economic** | | | | | | | | | | | | | | | | | | |
| Resource | | | | | | | | | | | X | | | | | | | |
| *Wind speed* | | X | | | X | X | | X | | | | X | X | | X | | | |
| *Irradiance* | X | | X | X | | | | | | | | | | | | | | |
| Material transport | X | X | X | | | X | | X | | | X | X | X | | X | | | |
| Connection cost | X | X | X | | | X | | | | | X | X | X | | X | | | X |
| Land Value | | X | | X | | X | | | X | | | | | | | | | |

| Group / Criteria / Sub-Criteria | Effat[38] | McKenna[18] | Szurek[36] | Sanchez [40] | Rodriguez [32] | Latinopoulos [29] | Tsoutsos [34] | Watson (wind) [35] | Watson (PV) [35] | Hoefer [28] | Holinger (Min) [1] | Holinger (Med) | Holinger (Max) | Samsatli [20] | Atici [30] | Hernandez [33] | Hernandez [33] |
|---|---|---|---|---|---|---|---|---|---|---|---|---|---|---|---|---|---|
| **Sociopolitical** | | | | | | | | | | | | | | | | | |
| Settlements | | X | X | | X | | | X | X | | X | X | X | X | | X | X |
| *Rural* | | | | | X | X | X | | | | X | X | X | | X | | |
| *Urban* | X | | | X | | X | X | | | | X | | | | X | | |
| Roads | X | | X | | X | X | X | | | X | | | | | | | |
| *Main* | | X | | | | | | | | | X | X | X | X | | | |
| *Secondary* | | X | | | | | | | | | X | X | X | | | | |
| Airports | | X | | X | X | X | X | | | | X | X | X | X | X | | |
| *Airports* | X | | | | | | | | | | | | | | | | |
| *Airfields* | | | | | | | | | | | | | | | | | |
| Power Lines | X | | X | | | | | X | | | X | X | X | | X | | |
| Historical Sites | X | | | | | X | X | X | X | X | | | | | | | |
| Railways | | X | X | | | | | | | | X | X | X | | X | | |
| Agriculture | | X | | X | | X | X | X | X | | | | | | | | |
| Industrial Areas | | X | | | | X | | | | | X | X | X | | | | |
| Mining sites | | | | | | X | X | | | X | | | | | X | | |
| Recreational Areas | | | | | | | | | | | | | | | | | |
| Camp sites | | | | | | | | | | | | | | | | | |
| Tourist sites | | | | | | X | X | | | X | | | | | | | |
| Radio Towers | | | | X | | | X | | | | | | | | X | | |
| Gas Network | | | | | | | | | | | | | | | | | |
| Power Plants | | | | | | | | | | | | | | | | | |
| **Physical** | | | | | | | | | | | | | | | | | |
| Water Bodies | X | | X | | X | | X | | | | X | X | X | | | X | X |
| *Lakes* | | | | | | | X | | | | X | X | X | | X | | |
| *Rivers* | | | | | | | X | | | | | | | X | X | | |
| Wetlands | | X | | | | | X | | | | | | | | | | |
| Slope | X | | X | X | X | X | X | X | X | X | X | X | X | X | X | | |
| Woodlands | | X | X | | | | X | | | | X | X | | | X | | |
| *Coniferous* | | | | | | | | | | | X | | | | | | |
| *Deciduous* | | | | | | | | | | | X | | | | | | |
| Vegetation | | | | | | | | | | | X | | | | | | |
| Land Instability | | | | | | | | | | | | | | | | | |
| *Earth quake* | | | | | | | | | | | | | | | X | | |
| *Flood plains* | | | | | | | | | | | | | | | | | |
| *Land slide* | | | | | | | | | | | | | | | | | |
| Elevation | | | | | | | | | | | X | X | X | | X | | |
| Soil Composition | | X | | | | | | | | X | | | | | | | |
| Aspect | | | X | | | | | | X | | | | | | | | |
| **Conservation** | | | | | | | | | | | | | | | | | |
| Protected Areas | X | | X | | | | X | | | | X | X | X | | X | X | X |
| *Parks* | | X | | | X | | X | | | | X | X | X | | | | |
| *Reserves* | | X | | | X | | | | | | | | | | | | |
| *Landscapes* | | X | | | X | X | | X | X | | | | | | | | |
| *Natural Monuments* | | | | | X | | | | | X | | | | | | | |
| *Wilderness* | | | | | | | X | | | | | | | | | | |
| Protected FFH | | X | X | | X | X | X | X | | X | | | | X | X | | |
| *Habitats* | X | X | | | X | | X | | | X | | | | | | X | X |
| *Biospheres* | | X | | | | | X | | | X | | | | | | | |
| **Economic** | | | | | | | | | | | | | | | | | |
| Resource | | | | | | | | | | | | | | | X | | |
| *Wind speed* | | | | X | X | X | | X | | X | | | | X | | | |
| *Irradiance* | | | | | | | | | X | | | | | | | X | X |
| Material transport | | | | X | | | | X | X | X | | | | X | | X | X |
| Connection cost | | | | X | | | | X | X | X | | | | X | | X | X |
| Land Value | | | | | | X | | | | | | | | | | | |

Appendix B: Prior Dataset Production Methods

For each of the created Prior datasets, a slightly unique procedure is followed. In all cases, however, the first step involved initially setting all pixels in the European study area to the "no indication" value, 254, while all values outside the study area are assigned the 'no data' value of 255. Then, for each predetermined edge, starting from the largest and proceeding in reverse order, all locations that have a criterion value less than the given edge are reassigned to the edge's index given by the table in Appendix C. When a criterion involves a simple value threshold, such as for elevations, the edge is applied directly to the value at each location. However, when the criterion involves a proximity threshold (such as the railway example provided earlier), edge containment is determined by buffering a 'raw' indication by the desired edge distance, followed by a rasterization identifying the pixels that lie in the buffered region. We now give more specific detail for each of the constructed Priors.

*B.I. Sociopolitical Priors*

Under the Sociopolitical motivation criteria group, 19 Prior datasets were created.

The **settlement proximity** Prior describes the distance from all settlement areas and is defined from locations indicated by the Corine Land Cover (CLC) [25] as "continuous urban fabric" (CLC-code 1.1.1) or as "Discontinuous urban fabric" (1.1.2). **Settlement urban proximity** gives distances from specifically dense urban settlements as indicated from features found directly in the European Commission's (EC) Urban Clusters [78] dataset, which was developed as part of the Geographic Information System of the Commission (GISCO) initiative. In this dataset, urban areas were derived from a population density dataset at 1 km resolution, where only contiguous regions possessing a minimum of 300 inhabitants per $km^2$ at each point and a total population above 1500 were identified as urban.

The **Airfield proximity** Prior describes distances from small airports and is produced by combining information from both CLC and GISCO's airport transportation network dataset [79]. In this case the EC's airport dataset is used to identify the coordinates of activate airports that do not report a large passenger throughput (i.e., less than 150,000 passengers per year, or otherwise unreported), which is then connected to the closest contiguous area larger than 1 $km^2$ from the CLC dataset being indicated as "Airports" (1.2.4). If no appropriate area is found within 2 km, a radius of 800 meters is assumed. Similarly, **Airport proximity** describes distances from large and commercial airports using the same procedure described for airfields; although in this case, annual passenger count must be greater than 150,000 and, if an appropriate area cannot be identified in the CLC dataset, a radius of 3 km is assumed.

**Roadway proximity**, **roadway main proximity** and **roadway secondary proximity** are all defined from a recent extract of the OpenStreetMap (OSM) [80] database and respectively describe distances from all roadways, major routes and highways, and local and state roadways. In order to remove unwanted features in the OSM dataset (such as footpaths and race tracks), roadway proximity accepts all routes and links classified under the "highway" key as "motorway", "trunk", "primary", "secondary", "tertiary", "service" and "unclassified". Meanwhile, roadway-main proximity only considers routes and links indicated as "motorway", "trunk" and "primary", while roadway-secondary proximity considers "secondary" and "tertiary".

The **agriculture proximity** Prior describes distances from all agricultural areas as indicated within the CLC agricultural group (2.X.X). Although not generally required by the LE literature sources, four additional agricultural-based Priors were created to reflect separate agricultural functionality, since this information was readily available from the CLC dataset.

**Agriculture arable proximity** is defined from CLC indications of "Non-irrigated arable land" (2.1.1), "permanently irrigated land" (2.1.2) or "rice fields" (2.1.3). **Agriculture permanent crop proximity** is taken from indications of "vineyards" (2.2.1), "fruit trees and berry plantations" (2.2.2) and "olive groves" (2.2.3). **Agriculture heterogeneous proximity** is taken from all CLC codes in the "heterogeneous agricultural areas" group (2.4.x). Finally, **agriculture pasture proximity** simply comes from indications of "pastures" (2.3.1).

Similar to the roadway priors, **railway proximity** and **power line proximity** describe distances from either the railway or the electricity grid network and are also taken from the OSM extract. Filtering is accomplished by selecting routes classified as "rail" under the "railway" key, or else as "line" under the "power" key.

Leisure proximity, camping proximity, and touristic proximity also describe distances from features as defined in the OSM dataset. **Leisure proximity** refers to expansive areas to which people may go for recreational or relaxation purposes and were taken as OSM features classified as "common" or "park" under the "leisure" key. **Camping proximity** describes distances from camp grounds and camp sites and are taken from OSM features classified as "camp_site" under the "tourism" key. **Touristic proximity** describes places of local and cultural importance and are taken from OSM features classified as "attraction" under the "tourism" key.

**Industrial proximity** refers to distances from areas or industrial and commercial activity outside of settlement areas, which are defined from areas indicated by CLC as "industrial or commercial units" (1.2.1).

As the last Prior in this group, **mining proximity** refers to distances from mining sites and are defined from areas indicated by the CLC as "mineral extraction sites" (1.3.1).

*B.II. Physical Priors*

Within the Physical motivation criteria group, 13 Prior datasets were created.

The **slope threshold** Prior describes areas with an average slope less than the chosen edges, as determined by calculating the pixel-wise gradient of the EuroDEM [81] dataset (another product of the GISCO initiative).

**Waterbody proximity** refers to distances from all open inland water bodies that are large enough to be classified as a single pixel in the Permanent Water Body [82] dataset developed by the Copernicus group (i.e., comprising the majority of a 20 m pixel) and can refer to lakes, large rivers, reservoirs and other such features. The water-body proximity does not distinguish between stagnant and running water bodies and, moreover, it does not pick up smaller streams and rivers; therefore, additional Prior datasets were created. **Lake proximity** describes distances from stagnant bodies of water as defined by the HydroLAKES [65] dataset developed by the World Wildlife Fund (WWF). **River proximity** refers to distances from probable routes of running water, including both large rivers and smaller streams, and is taken from GISCO's Hydrography [83] dataset. Routes in this dataset were seen to match well with the running bodies of water in the CLC dataset; however, only the route is provided, without any information on the body's width, so this Prior is best employed in addition to the waterbody proximity Prior in order to account for smaller rivers and streams. The **coast proximity** Prior describes the distances from coastlines as defined by CLC-indicated "sea and ocean" (5.2.3) areas.

Similar to the agricultural priors, the **woodland proximity** Prior describes distances from all woodland areas as indicated by the CLC dataset (3.1.X), but is also broken down into several sub-criterion priors according to the forest's primary composition, since this information was readily available from the CLC dataset. **Woodland-deciduous proximity**, **woodland-coniferous proximity** and **woodland-mixed proximity** refer, respectively, to areas indicated as "deciduous forest" (3.1.1), "coniferous forest" (3.1.2) and "mixed forest" (3.1.3).

**Wetland proximity** describes distances from wetlands, marshes and swamps as indicated from all areas indicated in the CLC dataset under the "wetlands" (4.X.X) category.

**Elevation-threshold** simply describes the areas that have an average elevation lower than the given edges as read directly by the EuroDEM dataset. While the Elevation-threshold Prior has the benefit of being much smaller in size than the original EuroDEM dataset (roughly 60 MB vs 20 GB), it does not derive any information beyond what is already communicated by the EuroDEM dataset and so it is suggested to continue using the original EuroDEM whenever possible.

As the final Prior in this group, the **sand proximity** describes distances from areas that are dominated by sand cover, as classified in the CLC dataset as "beaches, dunes, [and] sands" (3.3.1).

*B.III. Conservation Priors*

The Conservation motivation group is comprised of 8 Priors representing distances from features extracted from the World Database on Protected Areas (WDPA) [84] dataset.

**Protected habitat proximity** is defined from WDPA features with an International Union for the Conservation of Nature (IUNC) category of 'IV', or else having the word "habitat" within the area's English designation.

In a similar manner, **Protected wilderness proximity** is assigned as WPDA features with an IUNC category of 'Ib' or otherwise as having "wilderness" within the English designation.

**Protected biosphere proximity** and **Protected bird proximity** are given as WPDA features with "bio" or "bird," respectively, within the area's English designation.

**Protected park proximity**, **protected reserve proximity**, **protected natural-monument proximity** and **protected landscape proximity** refers to WDPA features with an IUNC category of 'II', 'Ia', 'III' or 'V', or as having an English designation containing the word "park," "reserve," "monument" or "landscape," respectively.

Due to the selection of features by both the IUNC category and contents of the English designation, there is significant overlap between these Prior datasets; however, this is to be expected as the original definitions of these areas are already overlapping.

*B.IV. Economic Priors*

In the final motivation group, economically-derived criteria, 6 Prior datasets were created, although two of these are simply aliases of previously generated Priors, as they can be used for these criteria as well.

**Access distance**, referring to distances away from the road network, is one such criterion which is an alias of the general roads proximity Prior. The same applies to **Connection distance**, describing distances away from the electricity network, is an alias of the power line proximity Prior.

For resource-related criteria like average annual wind speed or mean daily irradiance, several datasets are available, such as the DTU's Global Wind Atlas (GWA) [85] or the World Bank's Global Solar Atlas (GSA) [86]. These datasets are generally provided as a raster and, as such, can already be used directly in the methodology discussed in Section 3.2. Nevertheless, in a similar situation to the dataset used for the elevation Prior, Prior datasets have been constructed from these raw sources in order to expedite their use in LE analyses; however, the original sources should be preferred when available.

The **50m wind speed threshold** and **100m wind speed threshold** Priors were constructed directly by applying the chosen edges to the GWA's average wind speeds at 50 m and 100 m, respectively.

Similarly, the **ghi threshold** and **dni threshold** Priors were constructed from the direct application of the chosen edges onto the GSA's average daily global horizontal irradiance (ghi) and average daily direct normal irradiance (dni) datasets, respectively.

The GWA and GSA are both described at 1 km resolution across the whole world. The spatial interpolation of these datasets to the standard of 100 m was accomplished by using a cubic spline interpolation scheme. The resource-related datasets should be used with caution in LE analyses however, since, as expressed explicitly in the GWA's disclaimer, they are not intended for the direct micro-siting of RES systems.

## Appendix C: Prior Dataset Edges

| | Social & Political | | | | | | | | | | | | | | | | | | |
|---|---|---|---|---|---|---|---|---|---|---|---|---|---|---|---|---|---|---|---|
| index | Settlement proximity | Settlement urban proximity | Airport proximity | Airfield proximity | Roads proximity | Roads main proximity | Roads secondary proximity | Agriculture proximity | Agriculture arable proximity | Agriculture permanent crop proximity | Agriculture heterogeneous proximity | Agriculture pasture proximity | Railway proximity | Power line proximity | Leisure proximity | Camping proximity | Touristic proximity | Industrial proximity | Mining proximity |
| | km | km | km | km | km | km | km | km | km | km | km | km | km | km | km | km | km | km | km |
| 0 | 0 | 0 | 0 | 0 | 0 | 0 | 0 | 0 | 0 | 0 | 0 | 0 | 0 | 0 | 0 | 0 | 0 | 0 | 0 |
| 1 | 0.1 | 0.1 | 0.1 | 0.1 | 0.1 | 0.05 | 0.05 | 0.1 | 0.1 | 0.1 | 0.1 | 0.1 | 0.05 | 0.1 | 0.1 | 0.1 | 0.1 | 0.1 | 0.1 |
| 2 | 0.2 | 0.2 | 0.2 | 0.2 | 0.2 | 0.1 | 0.1 | 0.2 | 0.2 | 0.2 | 0.2 | 0.2 | 0.1 | 0.2 | 0.2 | 0.2 | 0.2 | 0.2 | 0.2 |
| 3 | 0.3 | 0.3 | 0.3 | 0.3 | 0.3 | 0.2 | 0.2 | 0.3 | 0.3 | 0.3 | 0.3 | 0.3 | 0.2 | 0.3 | 0.3 | 0.3 | 0.3 | 0.3 | 0.3 |
| 4 | 0.4 | 0.4 | 0.4 | 0.4 | 0.4 | 0.3 | 0.3 | 0.4 | 0.4 | 0.4 | 0.4 | 0.4 | 0.3 | 0.4 | 0.4 | 0.4 | 0.4 | 0.4 | 0.4 |
| 5 | 0.5 | 0.5 | 0.5 | 0.5 | 0.5 | 0.4 | 0.4 | 0.5 | 0.5 | 0.5 | 0.5 | 0.5 | 0.4 | 0.5 | 0.5 | 0.5 | 0.5 | 0.5 | 0.5 |
| 6 | 0.6 | 0.6 | 0.6 | 0.6 | 0.6 | 0.5 | 0.5 | 0.6 | 0.6 | 0.6 | 0.6 | 0.6 | 0.5 | 0.6 | 0.6 | 0.6 | 0.6 | 0.6 | 0.6 |
| 7 | 0.7 | 0.7 | 0.7 | 0.7 | 0.7 | 0.6 | 0.6 | 0.7 | 0.7 | 0.7 | 0.7 | 0.7 | 0.6 | 0.7 | 0.7 | 0.7 | 0.7 | 0.7 | 0.7 |
| 8 | 0.8 | 0.8 | 0.8 | 0.8 | 0.8 | 0.7 | 0.7 | 0.8 | 0.8 | 0.8 | 0.8 | 0.8 | 0.7 | 0.8 | 0.8 | 0.8 | 0.8 | 0.8 | 0.8 |
| 9 | 0.9 | 0.9 | 0.9 | 0.9 | 0.9 | 0.8 | 0.8 | 0.9 | 0.9 | 0.9 | 0.9 | 0.9 | 0.8 | 0.9 | 0.9 | 0.9 | 0.9 | 0.9 | 0.9 |
| 10 | 1 | 1 | 1 | 1 | 1 | 0.9 | 0.9 | 1 | 1 | 1 | 1 | 1 | 0.9 | 1 | 1 | 1 | 1 | 1 | 1 |
| 11 | 1.1 | 1.1 | 1.2 | 1.2 | 1.2 | 1 | 1 | 1.2 | 1.2 | 1.2 | 1.2 | 1.2 | 1 | 1.2 | 1.2 | 1.2 | 1.2 | 1.2 | 1.2 |
| 12 | 1.2 | 1.2 | 1.5 | 1.5 | 1.4 | 1.2 | 1.2 | 1.4 | 1.4 | 1.4 | 1.4 | 1.4 | 1.2 | 1.4 | 1.4 | 1.4 | 1.4 | 1.4 | 1.4 |
| 13 | 1.3 | 1.3 | 1.8 | 1.8 | 1.6 | 1.4 | 1.4 | 1.6 | 1.6 | 1.6 | 1.6 | 1.6 | 1.4 | 1.6 | 1.6 | 1.6 | 1.6 | 1.6 | 1.6 |
| 14 | 1.4 | 1.4 | 2 | 2 | 1.8 | 1.6 | 1.6 | 1.8 | 1.8 | 1.8 | 1.8 | 1.8 | 1.6 | 1.8 | 1.8 | 1.8 | 1.8 | 1.8 | 1.8 |
| 15 | 1.5 | 1.5 | 2.2 | 2.2 | 2 | 1.8 | 1.8 | 2 | 2 | 2 | 2 | 2 | 1.8 | 2 | 2 | 2 | 2 | 2 | 2 |
| 16 | 1.6 | 1.6 | 2.5 | 2.5 | 2.2 | 2 | 2 | 2.5 | 2.5 | 2.5 | 2.5 | 2.5 | 2 | 2.5 | 2.5 | 2.5 | 2.5 | 2.5 | 2.5 |
| 17 | 1.7 | 1.7 | 3 | 3 | 2.5 | 2.2 | 2.2 | 3 | 3 | 3 | 3 | 3 | 2.5 | 3 | 3 | 3 | 3 | 3 | 3 |
| 18 | 1.8 | 1.8 | 3.5 | 3.5 | 2.8 | 2.5 | 2.5 | 4 | 4 | 4 | 4 | 4 | 3 | 4 | 4 | 4 | 4 | 4 | 4 |
| 19 | 1.9 | 1.9 | 4 | 4 | 3 | 2.8 | 2.8 | 5 | 5 | 5 | 5 | 5 | 4 | 5 | 5 | 5 | 5 | 5 | 5 |
| 20 | 2 | 2 | 4.5 | 4.5 | 3.5 | 3 | 3 | | | | | | 5 | 6 | | | | | |
| 21 | 2.2 | 2.2 | 5 | 5 | 4 | 3.5 | 3.5 | | | | | | 6 | 7 | | | | | |
| 22 | 2.4 | 2.4 | 5.5 | 5.5 | 4.5 | 4 | 4 | | | | | | 7 | 8 | | | | | |
| 23 | 2.6 | 2.6 | 6 | 6 | 5 | 4.5 | 4.5 | | | | | | 8 | 9 | | | | | |
| 24 | 2.8 | 2.8 | 7 | 7 | 6 | 5 | 5 | | | | | | 9 | 10 | | | | | |
| 25 | 3 | 3 | 8 | 8 | 7 | 6 | 6 | | | | | | 10 | 12 | | | | | |
| 26 | 3.5 | 3.5 | 9 | 9 | 8 | 7 | 7 | | | | | | 12 | 14 | | | | | |
| 27 | 4 | 4 | 10 | 10 | 10 | 8 | 8 | | | | | | 14 | 16 | | | | | |
| 28 | 4.5 | 4.5 | 15 | 15 | 12 | 10 | 10 | | | | | | 16 | 18 | | | | | |
| 29 | 5 | 5 | | | 14 | 12 | 12 | | | | | | 18 | 20 | | | | | |
| 30 | 5.5 | 5.5 | | | 16 | 14 | 14 | | | | | | 20 | 25 | | | | | |
| 31 | 6 | 6 | | | 18 | 16 | 16 | | | | | | 25 | 30 | | | | | |
| 32 | 7 | 7 | | | 20 | 18 | 18 | | | | | | 30 | 35 | | | | | |
| 33 | 8 | 8 | | | | 20 | 20 | | | | | | 40 | 40 | | | | | |
| 34 | 9 | 9 | | | | 25 | 25 | | | | | | | 45 | | | | | |
| 35 | 10 | 10 | | | | 30 | 30 | | | | | | | 50 | | | | | |
| 36 | 15 | 15 | | | | 40 | 40 | | | | | | | | | | | | |
| 37 | 20 | 20 | | | | | | | | | | | | | | | | | |

| | Physical | | | | | | | | | | | |
|---|---|---|---|---|---|---|---|---|---|---|---|---|
| index | Slope threshold | Waterbody proximity | Lake proximity | River proximity | Ocean proximity | Woodland proximity | Woodland deciduous proximity | Woodland coniferous proximity | Woodland mixed proximity | Wetland proximity | Elevation threshold | Sand proximity |
| | deg | km | km | km | km | km | km | km | km | km | km | km |
| 0 | 0 | 0 | 0 | 0 | 0 | 0 | 0 | 0 | 0 | 0 | -1.0 | 0 |
| 1 | 0.5 | 0.1 | 0.1 | 0.05 | 0.1 | 0.1 | 0.1 | 0.1 | 0.1 | 0.1 | -0.9 | 0.05 |
| 2 | 1 | 0.2 | 0.2 | 0.1 | 0.2 | 0.2 | 0.2 | 0.2 | 0.2 | 0.2 | -0.8 | 0.1 |
| 3 | 1.5 | 0.3 | 0.3 | 0.2 | 0.3 | 0.3 | 0.3 | 0.3 | 0.3 | 0.3 | -0.7 | 0.2 |
| 4 | 2 | 0.4 | 0.4 | 0.3 | 0.4 | 0.4 | 0.4 | 0.4 | 0.4 | 0.4 | -0.6 | 0.3 |
| 5 | 2.5 | 0.5 | 0.5 | 0.4 | 0.5 | 0.5 | 0.5 | 0.5 | 0.5 | 0.5 | -0.5 | 0.4 |
| 6 | 3 | 0.6 | 0.6 | 0.5 | 0.6 | 0.6 | 0.6 | 0.6 | 0.6 | 0.6 | -0.4 | 0.5 |
| 7 | 3.5 | 0.7 | 0.7 | 0.6 | 0.7 | 0.7 | 0.7 | 0.7 | 0.7 | 0.7 | -0.3 | 0.6 |
| 8 | 4 | 0.8 | 0.8 | 0.7 | 0.8 | 0.8 | 0.8 | 0.8 | 0.8 | 0.8 | -0.2 | 0.7 |
| 9 | 4.5 | 0.9 | 0.9 | 0.8 | 0.9 | 0.9 | 0.9 | 0.9 | 0.9 | 0.9 | -0.1 | 0.8 |
| 10 | 5 | 1 | 1 | 0.9 | 1 | 1 | 1 | 1 | 1 | 1 | 0 | 0.9 |
| 11 | 5.5 | 1.2 | 1.2 | 1 | 1.2 | 1.2 | 1.2 | 1.2 | 1.2 | 1.2 | 0.1 | 1 |
| 12 | 6 | 1.4 | 1.4 | 1.2 | 1.4 | 1.4 | 1.4 | 1.4 | 1.4 | 1.4 | 0.2 | 1.2 |
| 13 | 6.5 | 1.6 | 1.6 | 1.4 | 1.6 | 1.6 | 1.6 | 1.6 | 1.6 | 1.6 | 0.3 | 1.4 |
| 14 | 7 | 1.8 | 1.8 | 1.6 | 1.8 | 1.8 | 1.8 | 1.8 | 1.8 | 1.8 | 0.4 | 1.6 |
| 15 | 7.5 | 2 | 2 | 1.8 | 2 | 2 | 2 | 2 | 2 | 2 | 0.5 | 1.8 |
| 16 | 8 | 2.5 | 2.5 | 2 | 2.5 | 2.5 | 2.5 | 2.5 | 2.5 | 2.5 | 0.6 | 2 |
| 17 | 8.5 | 3 | 3 | 2.5 | 3 | 3 | 3 | 3 | 3 | 3 | 0.7 | 2.5 |
| 18 | 9 | 4 | 4 | 3 | 4 | 4 | 4 | 4 | 4 | 4 | 0.8 | 3 |
| 19 | 9.5 | 5 | 5 | 4 | 5 | 5 | 5 | 5 | 5 | 5 | 0.9 | 4 |
| 20 | 10 | | | 5 | 10 | | | | | | 1 | 5 |
| 21 | 10.5 | | | | 15 | | | | | | 1.1 | |
| 22 | 11 | | | | 20 | | | | | | 1.2 | |
| 23 | 11.5 | | | | | | | | | | 1.3 | |
| 24 | 12 | | | | | | | | | | 1.4 | |
| 25 | 12.5 | | | | | | | | | | 1.5 | |
| 26 | 13 | | | | | | | | | | 1.6 | |
| 27 | 13.5 | | | | | | | | | | 1.7 | |
| 28 | 14 | | | | | | | | | | 1.8 | |
| 29 | 14.5 | | | | | | | | | | 1.9 | |
| 30 | 15 | | | | | | | | | | 2 | |
| 31 | 15.5 | | | | | | | | | | 2.1 | |
| 32 | 16 | | | | | | | | | | 2.2 | |
| 33 | 16.5 | | | | | | | | | | 2.3 | |
| 34 | 17 | | | | | | | | | | 2.4 | |
| 35 | 17.5 | | | | | | | | | | 2.5 | |
| 36 | 18 | | | | | | | | | | 2.6 | |
| 37 | 18.5 | | | | | | | | | | 2.7 | |
| 38 | 19 | | | | | | | | | | 2.8 | |
| 39 | 19.5 | | | | | | | | | | 2.9 | |
| 40 | 20 | | | | | | | | | | 3 | |
| ... | ... | | | | | | | | | | | |
| 55 | 27.5 | | | | | | | | | | | |
| 56 | 28 | | | | | | | | | | | |
| 57 | 28.5 | | | | | | | | | | | |
| 58 | 29 | | | | | | | | | | | |
| 59 | 29.5 | | | | | | | | | | | |
| 60 | 30 | | | | | | | | | | | |

| | Conservation | | | | | | | | Economical | | | | | |
|---|---|---|---|---|---|---|---|---|---|---|---|---|---|---|
| index | Protected habitat proximity | Protected bird proximity | Protected biosphere proximity | Protected wilderness proximity | Protected landscape proximity | Protected reserve proximity | Protected park proximity | Protected natural monument proximity | Windspeed 50m threshold | Windspeed 100m threshold | DNI threshold | GHI threshold | Access distance | Connection distance |
| | km | km | km | deg | km | km | km | km | m/s | m/s | $\frac{kWh}{m^2 day}$ | $\frac{kWh}{m^2 day}$ | km | km |
| 0 | 0 | 0 | 0 | 0 | 0 | 0 | 0 | 0 | 0 | 0 | 0 | 0 | 0 | 0 |
| 1 | 0.1 | 0.1 | 0.1 | 0.1 | 0.1 | 0.1 | 0.1 | 0.1 | 0.25 | 0.25 | 0.25 | 0.25 | 0.1 | 0.1 |
| 2 | 0.2 | 0.2 | 0.2 | 0.2 | 0.2 | 0.2 | 0.2 | 0.2 | 0.5 | 0.5 | 0.5 | 0.5 | 0.2 | 0.2 |
| 3 | 0.3 | 0.3 | 0.3 | 0.3 | 0.3 | 0.3 | 0.3 | 0.3 | 0.75 | 0.75 | 0.75 | 0.75 | 0.3 | 0.3 |
| 4 | 0.4 | 0.4 | 0.4 | 0.4 | 0.4 | 0.4 | 0.4 | 0.4 | 1 | 1 | 1 | 1 | 0.4 | 0.4 |
| 5 | 0.5 | 0.5 | 0.5 | 0.5 | 0.5 | 0.5 | 0.5 | 0.5 | 1.25 | 1.25 | 1.25 | 1.25 | 0.5 | 0.5 |
| 6 | 0.6 | 0.6 | 0.6 | 0.6 | 0.6 | 0.6 | 0.6 | 0.6 | 1.5 | 1.5 | 1.5 | 1.5 | 0.6 | 0.6 |
| 7 | 0.7 | 0.7 | 0.7 | 0.7 | 0.7 | 0.7 | 0.7 | 0.7 | 1.75 | 1.75 | 1.75 | 1.75 | 0.7 | 0.7 |
| 8 | 0.8 | 0.8 | 0.8 | 0.8 | 0.8 | 0.8 | 0.8 | 0.8 | 2 | 2 | 2 | 2 | 0.8 | 0.8 |
| 9 | 0.9 | 0.9 | 0.9 | 0.9 | 0.9 | 0.9 | 0.9 | 0.9 | 2.25 | 2.25 | 2.25 | 2.25 | 0.9 | 0.9 |
| 10 | 1 | 1 | 1 | 1 | 1 | 1 | 1 | 1 | 2.5 | 2.5 | 2.5 | 2.5 | 1 | 1 |
| 11 | 1.2 | 1.2 | 1.2 | 1.2 | 1.2 | 1.2 | 1.2 | 1.2 | 2.75 | 2.75 | 2.75 | 2.75 | 1.2 | 1.2 |
| 12 | 1.4 | 1.4 | 1.4 | 1.4 | 1.4 | 1.4 | 1.4 | 1.4 | 3 | 3 | 3 | 3 | 1.4 | 1.4 |
| 13 | 1.6 | 1.6 | 1.6 | 1.6 | 1.6 | 1.6 | 1.6 | 1.6 | 3.25 | 3.25 | 3.25 | 3.25 | 1.6 | 1.6 |
| 14 | 1.8 | 1.8 | 1.8 | 1.8 | 1.8 | 1.8 | 1.8 | 1.8 | 3.5 | 3.5 | 3.5 | 3.5 | 1.8 | 1.8 |
| 15 | 2 | 2 | 2 | 2 | 2 | 2 | 2 | 2 | 3.75 | 3.75 | 3.75 | 3.75 | 2 | 2 |
| 16 | 2.5 | 2.5 | 2.5 | 2.5 | 2.5 | 2.5 | 2.5 | 2.5 | 4 | 4 | 4 | 4 | 2.2 | 2.5 |
| 17 | 3 | 3 | 3 | 3 | 3 | 3 | 3 | 3 | 4.25 | 4.25 | 4.25 | 4.25 | 2.5 | 3 |
| 18 | 4 | 4 | 4 | 4 | 4 | 4 | 4 | 4 | 4.5 | 4.5 | 4.5 | 4.5 | 2.8 | 4 |
| 19 | 5 | 5 | 5 | 5 | 5 | 5 | 5 | 5 | 4.75 | 4.75 | 4.75 | 4.75 | 3 | 5 |
| 20 | | | | | | | | | 5 | 5 | 5 | 5 | 3.5 | 6 |
| 21 | | | | | | | | | 5.25 | 5.25 | 5.25 | 5.25 | 4 | 7 |
| 22 | | | | | | | | | 5.5 | 5.5 | 5.5 | 5.5 | 4.5 | 8 |
| 23 | | | | | | | | | 5.75 | 5.75 | 5.75 | 5.75 | 5 | 9 |
| 24 | | | | | | | | | 6 | 6 | 6 | 6 | 6 | 10 |
| 25 | | | | | | | | | 6.25 | 6.25 | 6.25 | 6.25 | 7 | 12 |
| 26 | | | | | | | | | 6.5 | 6.5 | 6.5 | 6.5 | 8 | 14 |
| 27 | | | | | | | | | 6.75 | 6.75 | 6.75 | 6.75 | 10 | 16 |
| 28 | | | | | | | | | 7 | 7 | 7 | 7 | 12 | 18 |
| 29 | | | | | | | | | 7.25 | 7.25 | 7.25 | 7.25 | 14 | 20 |
| 30 | | | | | | | | | 7.5 | 7.5 | 7.5 | 7.5 | 16 | 25 |
| 31 | | | | | | | | | 7.75 | 7.75 | 7.75 | 7.75 | 18 | 30 |
| 32 | | | | | | | | | 8 | 8 | 8 | 8 | 20 | 35 |
| 33 | | | | | | | | | 8.25 | 8.25 | 8.25 | 8.25 | | 40 |
| 34 | | | | | | | | | 8.5 | 8.5 | 8.5 | 8.5 | | 45 |
| 35 | | | | | | | | | 8.75 | 8.75 | 8.75 | 8.75 | | 50 |
| 36 | | | | | | | | | 9 | 9 | 9 | 9 | | |
| 37 | | | | | | | | | 9.25 | 9.25 | 9.25 | 9.25 | | |
| 38 | | | | | | | | | 9.5 | 9.5 | 9.5 | 9.5 | | |
| 39 | | | | | | | | | 9.75 | 9.75 | 9.75 | 9.75 | | |
| 40 | | | | | | | | | 10 | 10 | 10 | 10 | | |
| ... | | | | | | | | | ... | ... | ... | ... | | |
| 75 | | | | | | | | | 18.8 | 18.8 | 18.8 | 18.8 | | |
| 76 | | | | | | | | | 19 | 19 | 19 | 19 | | |
| 77 | | | | | | | | | 19.3 | 19.3 | 19.3 | 19.3 | | |
| 78 | | | | | | | | | 19.5 | 19.5 | 19.5 | 19.5 | | |
| 79 | | | | | | | | | 19.8 | 19.8 | 19.8 | 19.8 | | |
| 80 | | | | | | | | | 20 | 20 | 20 | 20 | | |